\documentclass[sigconf,dvipsnames]{acmart}

\AtBeginDocument{%
  }

\copyrightyear{2026}
\acmYear{2026}
\setcopyright{cc}
\setcctype{by}
\acmConference[CHI '26]{Proceedings of the 2026 CHI Conference on Human Factors in Computing Systems}{April 13--17, 2026}{Barcelona, Spain}
\acmBooktitle{Proceedings of the 2026 CHI Conference on Human Factors in Computing Systems (CHI '26), April 13--17, 2026, Barcelona, Spain}
\acmPrice{}
\acmDOI{10.1145/3772318.3790415}
\acmISBN{979-8-4007-2278-3/2026/04}

\PassOptionsToPackage{dvipsnames}{xcolor}
\usepackage[dvipsnames]{xcolor}
\usepackage{booktabs, array, graphicx}
\usepackage{pifont}
\usepackage{multirow}
\usepackage{multicol}
\usepackage{float}

\newcommand{\rothead}[2][60]{%
  \rotatebox[origin=bl]{#1}{\makebox[0pt][l]{\centering #2}}%
}
\begin{document}

\title{Governance of AI-Generated Content: A Case Study on Social Media Platforms}

\author{Lan Gao}
\affiliation{%
  \institution{University of Chicago}
  \city{Chicago}
  \country{USA}}
\email{langao@uchicago.edu}
  
\author{Abani Ahmed}
\affiliation{%
  \institution{University of Chicago}
  \city{Chicago}
  \country{USA}}
\email{abani@uchicago.edu}

\author{Oscar Chen}
\affiliation{%
  \institution{University of Chicago}
  \city{Chicago}
  \country{USA}}
\email{oscarc@uchicago.edu}

\author{Margaux Reyl}
\affiliation{%
  \institution{University of Chicago}
  \city{Chicago}
  \country{USA}}
\email{margauxreyl@uchicago.edu}
  
\author{Zayna Cheema}
\affiliation{%
  \institution{University of Chicago}
  \city{Chicago}
  \country{USA}}
\email{zaynacheema@uchicago.edu}

\author{Nick Feamster}
\affiliation{%
  \institution{University of Chicago}
  \city{Chicago}
  \country{USA}}
\email{feamster@uchicago.edu}

\author{Chenhao Tan}
\affiliation{%
  \institution{University of Chicago}
  \city{Chicago}
  \country{USA}}
\email{chenhao@chenhaot.com}
  
\author{Kurt Thomas}
\affiliation{%
  \institution{Google}
  \city{Mountain View}
  \country{USA}}
\email{kurtthomas@google.com}
  
\author{Marshini Chetty}
\affiliation{%
  \institution{University of Chicago}
  \city{Chicago}
  \country{USA}}
\email{marshini@uchicago.edu}

\renewcommand{\shortauthors}{Gao et al.}

\begin{abstract} 
Online platforms are seeing increasing amounts of AI-generated content---text and other forms of media that are made or co-created with generative AI. This trend suggests platforms may need to establish governance frameworks, including policies and enforcement strategies for how users create, post, share, and engage with such content to encourage responsible use. We investigate the governance of AI-generated content across 40 popular social media platforms. Just over two-thirds explicitly describe governance of AI-generated content spanning six themes. Most platforms focus on moderating AI-generated content that violates established content rules and discloses AI-generated content. Fewer platforms---those that are focused on creativity and knowledge-sharing---address other issues such as ownership and monetization. Based on these findings, we suggest stakeholders and policymakers develop more direct, comprehensive, and forward-looking AI-generated content governance, as well as tools and education for users about the use of such content. 
\end{abstract}







\maketitle

\section{Introduction}

With the rise of consumer-facing generative AI products (e.g., ChatGPT),
AI-generated content\footnote{We note that the term \textit{AI-generated
content} in this paper does not refer only to fully AI-generated materials. It
also encompasses content partially created with AI involvement or co-created by
humans and AI, such as \textit{AI-assisted content} and \textit{AI synthetic
media}. As no unified term exists for all such content, we adopt the common term
AI-generated content throughout.} has flooded online spaces~\cite{wei2024pixiv,
sun2025aigt, matatov2024subreddits, liang2025widespread}, reshaping digital
communication and media consumption, while raising new concerns about content
trustworthiness. Generative AI products lower the threshold to produce and
spread content at scale, including misinformation and non-consensual
deepfakes~\cite{gibson2025analyzing,zhou2023synthetic}. AI-generated content
can also polarize public speech and weaken perceived
trust~\cite{moller2025impact, chen2025synthetic, rae2024effects}. Moreover,
ongoing ambiguities around authorship and copyright of AI-generated content
further complicate accountability for such content~\cite{lima2025public,
kyi2025governance, samuelson2023generative}. These concerns highlight an urgent
need for effective \textit{governance
of AI-generated content} in online spaces. In this paper, we specifically concentrate on platform-level governance---that is, how online platforms establish policies, principles, and mechanisms to manage the creation, presentation, and distribution of AI-generated
content in ways that align with public expectations and regulatory requirements. While other parties, such as users, regulators, and AI service providers, also contribute to the governance of AI-generated content, we focus on the responsibilities and actions undertaken by platforms.

On platforms hosting user-generated content, the effects of generative AI are even more critical, as generative AI has been increasingly
adopted by users and creators~\cite{hua2024generative, lyu2024youtube,
kim2024unlocking}. Many platforms have also introduced their own custom
generative AI tools for content creation (e.g., Esty's ``Seller Handbook''
tool\footnote{\url{https://www.etsy.com/seller-handbook/article/1402347260856}} and
Tiktok's ``AI Alive''\footnote
{\url{https://newsroom.tiktok.com/en-us/introducing-tiktok-ai-alive}}). These changes
have shaped user dynamics, content ecology, and platform
infrastructure~\cite{hua2024generative}. Together with the potential concerns
associated with the trustworthiness of AI-generated content, platforms face
significant challenges in maintaining reliable and safe communities through effective governance approaches. 

Previous works have explored the
growth of generative AI used in user-generated content and its impact on online
platforms (e.g.,~\cite{hua2024generative, wei2024pixiv,
matatov2024subreddits, lyu2024youtube}), and evaluated governance
approaches such as labeling AI-generated content~\cite{burrus2024unmasking,
gamage2025labeling, jung2025ai, Wittenberg2024Labeling, epstein2023label}. Yet,
few studies offer a systematic understanding of how platform governance has shifted to respond to
the rise of AI-generated content and users employing generative AI tools. Notably, Lloyd et al. investigated the
user-driven, bottom-up governance of AI-generated content in
Reddit~\cite{lloyd2023there, lloyd2025ai}; but our
understanding of platform-driven, top-down governance of such content
remains limited. There have also been few regulations addressing the publishing and distribution of AI-generated content on the Internet. To inform regulations, platform policies, and safety measures for generative AI use in content creation, it is important to fill these gaps in our understanding. 

In this paper, we study the governance of AI-generated content on social
media---a type of platform hosting user-generated content that is used to
create, modify, share, and discuss Internet content~\cite{kietzmann2011social}.
These platforms offer a critical case study for our work because they are the largest digital medium where people interact with
content, other users~\cite{pew2024socialmedia}, content creators,
and advertisers. They also have to manage complex platform governance actions, from taking down harmful user-generated content to facilitating content monetization
~\cite{singhal2023sok, ma2022fairness}. In addition, these platforms have to deal with a high volume of content, multiple content types, and unpredictable user
behaviors---unlike eCommerce and news sites, which also host user-generated content---which complicates governance over content in general.

To meet our study goals, we systematically scraped web pages from the 40 most popular social media platforms and curated a dataset of pages containing platform responses on AI-generated content in their community. Through this process, we found that about two-thirds of platforms explicitly mention how they govern AI-generated content, with the rest relying on existing content moderation policies to cover such content. Platforms in our study describe six types of governance actions on AI-generated content (See Table \ref{tab:govsum}), including: 1) extending their existing content moderation policies to AI-generated content, 2) disclosing and labeling AI-generated content, 3) enacting AI-specific restrictions on AI-generated content, 4) constraining monetization of AI-generated content, 5) safeguarding output generation and distribution from platforms' integrated AI tools, and 6) providing resources for users to better learn about and control AI-generated content in their feeds. Based on our findings, we suggest that social media and other online platforms incorporate more explicit and comprehensive governance frameworks on AI-generated content. We also recommend that researchers and platform designers develop improved tools and resources for users to create, share, and interact with AI-generated content more responsibly and with more autonomy.

We make the following contributions to the HCI community: 
\begin{itemize}
    \item We created an annotated dataset of the web pages of platform responses to AI-generated content on 40 of the most popular social media platforms.\footnote{The dataset is available at \url{https://github.com/UChicagoAIRLab/aigc-governance-social-media}.}
    \item We provided a baseline of social media platforms' current governance actions on AI-generated content.
    \item We outlined a forward-looking agenda for the future development of governance of AI-generated content in the online space.
\end{itemize}


\section{Background and Related Work}

We first survey the regulatory landscape of AI-generated content, its impact on
online communities, and broader platform governance approaches. We then describe
how our work extends prior research on content moderation to understand how
platforms specifically govern AI-generated content.

\subsection{Background: Regulating AI-Generated Content in Online Spaces}

The invention of multi-modal generative neural networks in the 2010s, such as
generative adversarial networks (GANs)~\cite{goodfellow2014generative}, enabled
the creation of deepfakes---AI-generated realistic media depicting people, which
were soon exploited for disinformation, scams, and non-consensual media
online~\cite{westerlund2019emergence, al2023impact, ma2025social,
han2025characterizing}. Regulatory attention on AI-generated content originated
from the distribution of these illegal deepfakes that were highly damaging to
individuals' mental health and reputation. For example, the United States (US)
has enacted multiple state laws to mitigate harmful deepfake distribution since
2019 (e.g., Texas S.B. 751 on political deepfakes~\cite{Texas_SB751_2019} and
S.B. 1361 on pornographic deepfakes~\cite{Texas_SB1361_2023}). Recently, the US Federal government also passed a "Take It Down Act" to impose criminal penalties on individuals publishing non-consensual deepfakes online~\cite{US_TakeItDown_2025}. After this act passed, MrDeepFakes, the largest AI-generated pornography site, shut down in May 2025 owing to the loss of a critical service provider~\cite{CBS_MrDeepfakes_Shutdown_2025}.

Since the rise of consumer-facing generative AI products following ChatGPT's
release in 2022, the volume of AI-generated content is increasing online, and
the potential threats are no longer restricted to harmful deepfakes. Beyond laws addressing the distribution of harmful deepfakes, additional regulations targeting generative AI models and consumer-facing generative AI services were established to require that these tools adhere to safety, transparency, and accountability rules~\cite{luna2024navigating}. For instance, both the European Union
(EU) AI Act~\cite{EU_AIAct_Art50_2024} and China's Generative AI
Measures~\cite{China_GenerativeAI_Interim_2023} prohibit services tailored to
generate harmful content, and require service providers to add conspicuous and
machine-readable markers on AI-generated output. Broader regulations on
AI-generated content distribution on the Internet, however, remain rare. To date, China is the only country that has enacted nationwide rules---Measures for Labeling AI-Generated Content~\cite{CLT2025AILabeling}---that strictly require labeling any AI-generated content of any modality and its distribution within any public online space. Furthermore, Luna et al. researched AI governance guidelines in six countries and found that, at the time of study, only the US and the EU had established principles for managing the distribution of AI-generated content~\cite{luna2024navigating}. 

With the lack of related regulation guidance until now, the governance of online AI-generated content distribution outside harmful deepfakes is shaped by industrial standards and dominated by platforms themselves. Disclosing AI-generated content through labeling is the most widely known rising governance approach to AI-generated content. Industry has pushed provenance and metadata approaches that include machine-readable content origination information to support identifying AI-generated content (e.g., the ``Content Credentials''~\cite{C2PA2025, feng2023examining}). In academia, a growing body of research has focused on improving and evaluating the design of AI disclosure labels, finding that labels can both improve and compromise user engagement, trustworthiness, and recognition of misleading content in the use of online communities~\cite{burrus2024unmasking, gamage2025labeling, jung2025ai, Wittenberg2024Labeling, epstein2023label}. Most closely related to our own work, Lloyd et al. mapped community-level, bottom-up governance responses to AI-generated content on Reddit communities~\cite{lloyd2023there, lloyd2025ai}. This work, however, only covers community-driven policymaking on AI-generated content on one platform.

To inform the future directions of AI-generated content governance, our work extends existing studies on understanding and developing governance actions to provide a cross-platform, holistic review of the most popular social media platforms' policies and enforcement on AI-generated content.

\subsection{The Impact of AI-Generated Content in Online Communities}
Researchers have studied the impact of the increasing volume of AI-generated content on a broad range of online communities, where diverse groups of users interact around user-generated content. A recent work found a sharp increase in AI-generated textual posts on long-form text-based platforms such as Medium, Quora, and Reddit after the launch of ChatGPT \cite{sun2025aigt}. On platforms that are dominated by image and visual arts, researchers also found a visible spike in AI-generated artwork~\cite{matatov2024subreddits, wei2024pixiv}. A few recent prior studies also reveal the increasing adoption of generative AI tools by video creators in social media \cite{lyu2024youtube, kim2024unlocking}.

The presence of AI-generated content in user-generated content has reshaped user dynamics and public speech in online communities. On a positive note, researchers have pointed out that generative AI lowers barriers for new creators to enter a community \cite{matatov2024subreddits, wei2024pixiv}, encourages creators' in-depth participation \cite{moller2025impact}, and incentivizes more active or positive discussions among general users~\cite{chen2025synthetic, gao2025does}. A research study also found that user-generated online comments could be less polarized through human-AI co-creation~\cite{shahid2024examining}.  However, these benefits may also come with tradeoffs. Through an experimental study, Moller et al. revealed that generative AI tools can increase user participation but also diminish perceived authenticity and conversational quality \cite{moller2025impact}. Some prior works also reported decreased trustworthiness and quality of online content with the introduction of generative AI~\cite{burtch2023consequences, rae2024effects}. Researchers have also studied how AI-generated content can be distributed and compromise community safety through AI-generated misinformation~\cite{dmonte2024electionclaims, diresta2024spammers, kreps2022all, al2023impact}, AI impersonation \cite{mink2023deepphish}, and non-consensual deepfakes~\cite{ma2025social}. 

In light of this thread of work, it is apparent that the governance of AI-generated content is increasingly necessary, as platform policies could directly affect user participation, trust, and community dynamics. While prior work has widely investigated the rising social impacts of AI-generated content in online communities, less attention has been paid to how platforms establish governance mechanisms to address such impacts. Using social media as a case study, our work examines how online platforms have evolved their governance approaches to respond to the increase of AI-generated content in their communities.

\subsection{Content Moderation and Platform Governance in Online Communities}
To mitigate inappropriate content in their communities, online platforms need to decide whether to publish, remove, or flag user-generated content. This approach is usually referred to as content moderation \cite{kiesler2012regulating}. Traditionally, content moderation research has largely investigated the effects of moderation on community behaviors~\cite{chandrasekharan2017you, jhaver2019does} and user perceptions of content moderation mechanisms~\cite{ma2023users, myerswest2018censored, jhaver2019did, lyons2022s, vaccaro2020end}. A nascent line of research has also investigated the user rules and policies underpinning content moderation established in online communities~\cite{jiang2020characterizing, schaffner2024community, singhal2023sok, buckley2022censorship, fiesler2018reddit, fiesler2016reality}. Content moderation policies play a critical role in informing users of a community's rules and enforcement practices, yet cross-platform studies reveal that they differ across platforms and that there is a lack of a unified framework for such policymaking~\cite{singhal2023sok, schaffner2024community}.

With the development of generative AI like large language models (LLMs), both researchers and platforms have increasingly focused on how these systems can support content moderation practices~\cite{kolla2024llm,thomas2025supporting,communitynotes_x_api_overview}. In comparison, less research has explored how platforms moderate AI-generated content that violates existing community guidelines~\cite{fisher2024moderating}. Existing research mainly focuses on investigating the performance and failure modes of current moderation approaches on AI-generated content that is considered harmful in some way. For example, Mink et al. found that human reviewers moderating deepfake LinkedIn profiles are significantly biased based on the profile's demographic identity \cite{mink2024s}. Li et al. conducted an audit study on Twitter/X's platform user reports,  which showed that these are inadequate in reporting non-consensual deepfakes~\cite{qiwei2024deepfakes}.

Platform actions on AI-generated content are situated within a broader framework of platform governance than content moderation alone. As defined by Gorwa \cite{gorwa2019platform}, platform governance focuses on any platform actions and enforcement concerning online participation, typically involving three parties: platform companies themselves, platform users, and the governments of the jurisdictions in which the platform operates. In online communities, platform governance extends beyond the moderation of content to socio-technical arrangements like managing creator economies (how content creators make content and monetize it), educating users, and designing technical affordances that shape content ecology \cite{ma2022fairness, obrien2025dating, matias2018civilservant}. As revealed in prior works, the handling of emerging threats in online communities is usually a mixture of moderation and other efforts---all fall into the framework of platform governance. 

While prior works have examined how well harmful deepfakes are moderated~\cite{mink2024s, qiwei2024deepfakes}, there is a lack of understanding of the landscape of how platforms rule and act on AI-generated content, particularly after the introduction of consumer-facing generative AI products. We extend work on traditional content moderation to understanding social media platforms' broader governance actions on AI-generated content.
\section{Method}
We investigated 40 popular social media platforms to systematically understand
how they respond to and govern AI-generated content. Our data collection and
analysis focused on platform web pages that present policies, actions, and
stances on platform governance and community development. This includes Terms
of Service and Community Guidelines, as well as informal documentation such as
the Newsroom and Official Blog sections on these sites, as applicable. 

\begin{table*}[htbp]
\centering
\caption{The 40 social media platforms selected for our study in the format of ``platform (homepage URL)'', ordered alphabetically.}
\label{tab:platformlist}
\renewcommand{\arraystretch}{1.1}
\resizebox{0.99\linewidth}{!}{
\begin{tabular}{p{\textwidth}}
\toprule
\multicolumn{1}{c}{\textbf{Platform (homepage URL)}} \\ \midrule
Behance (\textit{behance.net}), Bluesky (\textit{bsky.app}), Dailymotion (\textit{dailymotion.com}), DeviantArt (\textit{deviantart.com}), 
Eporner (\textit{eporner.com}), \\Erome (\textit{erome.com}), Facebook (\textit{facebook.com}), Fandom (\textit{fandom.com}), 
Flickr (\textit{flickr.com}), Goodreads (\textit{goodreads.com}), Imgur (\textit{imgur.com}), \\Instagram (\textit{instagram.com}), 
LinkedIn (\textit{linkedin.com}), LiveJournal (\textit{livejournal.com}), Medium (\textit{medium.com}), OnlyFans (\textit{onlyfans.com}), 
\\Patreon (\textit{patreon.com}), Pinterest (\textit{pinterest.com}), Pixabay (\textit{pixabay.com}), Pixiv (\textit{pixiv.net}), 
Pornhub (\textit{pornhub.com}), Quora (\textit{quora.com}), \\Reddit (\textit{reddit.com}), ResearchGate (\textit{researchgate.net}), 
Roblox (\textit{roblox.com}), SoundCloud (\textit{soundcloud.com}), StackOverflow (\textit{stackoverflow.com}), \\SteamCommunity (\textit{steamcommunity.com}), 
Substack (\textit{substack.com}), Threads (\textit{threads.net}), TikTok (\textit{tiktok.com}), TradingView (\textit{tradingview.com}), \\
Tumblr (\textit{tumblr.com}), Twitch (\textit{twitch.tv}), Twitter/X (\textit{twitter.com}), Vimeo (\textit{vimeo.com}), 
XHamster (\textit{xhamster.com}), XVideos (\textit{xvideos.com}), \\Yelp (\textit{yelp.com}), YouTube (\textit{youtube.com}) \\
\bottomrule
\end{tabular}}
\end{table*}

\begin{table*}
    \centering
    \caption{Keyword lists for seed link collection and scraper}
    \label{tab:keywords}
    \resizebox{0.99\linewidth}{!}{
    \renewcommand{\arraystretch}{1.1}
    \begin{tabular}{|p{0.27\textwidth}|p{0.65\textwidth}|}
    \hline
        Keywords for seed link collection (used for search engine)*  & `generative ai', `genai', `large language model', `llm', `ai generated', `ai assisted', `ai made', `ai content', `ai media', `synthetic content', `deepfake' \\
    \hline
        Keywords for scraping and passages extraction (input in the scraper)** & `ai', `a.i.', `ai-', `-ai', `artificial intelligen', `genai', `llm', `large language model', `deepfake', `deep fake', `synthetic'\\
    \hline
    \multicolumn{2}{l}{\vspace{-0.5em}\footnotesize *When searching for seed links, we also used alternate forms of each keyword, such as `ai-generated' and `deep fake'.}\\
    \multicolumn{2}{l}{\footnotesize **Case insensitive. Applied wildcard to both sides of the keywords, except the keyword `ai'.}
    \end{tabular}}
\end{table*}

\subsection{Platform Selection}

We used Kietzmann et al.'s definition of social media: online platforms that enable users to create, modify, share, and discuss internet content~\cite{kietzmann2011social} as adopted by prior researchers (e.g.,~\cite{schaffner2022understanding}). We selected representative social media platforms for our analysis based on their popularity, following the common practice in prior cross-platform studies of social media (e.g., \cite{schaffner2022understanding, pater2016characterizations, almansoori2025can}). We first selected 20 social media platforms from Schaffner et al.'s study of content moderation rules on 43 platforms with user-generated content~\cite{schaffner2024community}. We also selected 20 additional platforms from the top 1000 websites from the Tranco List obtained on March 22nd, 2025.\footnote{Tranco List is a network traffic ranking resistant to manipulation and widely adopted in prior research as a measure of website popularity (e.g., \cite{schaffner2022understanding, schaffner2024community, habib2022okay}). This version of Tranco List can be accessed at: \url{https://tranco-list.eu/list/24879}} When selecting the 40 platforms (see Table \ref{tab:platformlist}), two research team members visited each website, checked if it provided English services, explored its core features and ``About'' pages, and assessed whether it fit our social media definition. The research team met regularly to discuss and resolve conflicts regarding the inclusion or exclusion of platforms. The final list of platforms is not meant to be exhaustive but represents a diverse spectrum of social media types, including but not limited to: large multi-purpose platforms such as Facebook and Twitter/X; social networking sites such as LinkedIn; video-focused platforms such as TikTok and Vimeo; creativity communities for writing, art, and music, such as Medium, DeviantArt, and SoundCloud; knowledge-sharing communities such as Stack Overflow and Quora; and platforms for explicit content, such as Pornhub and XHamster. The platforms vary in platform sizes, user bases, and content domains, allowing us to capture both common practices on AI-generated content from a rich sample of representative social media platforms, as well as variations linked to platform-specific attributes.

\subsection{Data Collection}
We performed data collection from April to June 2025. While many prior works on online platform document analysis rely on manual search to collect relevant documents (e.g., \cite{gao2025cannot, obrien2025dating}), for an exhaustive data collection, we used a web scraper to collect platforms' statements on AI-generated content spanning different types of web pages. We used the open source web scraper developed by Schaffner et al. to collect content moderation policies~\cite{schaffner2024community} for formal data collection. The scraper initializes its workflow from a set of \textit{seed links} provided as input and scrapes and parses the HTML from pages containing any terms from a predefined \textit{topic-wise keyword list}. 

We input two types of seed links: 1) \textit{canonical links}, include each platform's key policy pages (e.g., Terms of Service, Acceptable Use Policies, Community Guidelines), Support and Help Center, Newsroom, Announcement, and Official Blog pages, as well as links to AI-specific policy pages if they existed; and 2) pages addressing AI-generated content in the community. To find web pages addressing AI-generated content, we use a keyword list with both generative AI-related words and AI-generated content-related words (See Table \ref{tab:keywords}). We searched for these keywords and each platform name using Google search and checked at least the first three pages to identify related platform web pages. If applicable, we also searched keywords using the platform's internal search engines in Support/Help Center, Newsroom, Announcements, and Official Blog. Two researchers checked the completeness of our seed link collection for each platform. The seed links set we generated contains an average of 7.6 links per platform. 

As we found platforms use a variety of terms when referring to AI-generated content, we developed another keyword list that contains broader, AI-related keywords (See Table \ref{tab:keywords}) as the topic-wise keyword list input for the scraper. This enabled us to collect a wide range of pages where AI is mentioned, after which we could filter out the pages we wanted around AI-generated content. We piloted and tested the scraper on each platform using one or two seed links to ensure it was robust, then performed formal data collection using the scraper in May and June 2025. We eventually scraped 2518 pages through this process.

\subsection{Data Cleaning and Analysis}
Before the data analysis, we first ran the passage extraction module of the scraper on our collected pages~\cite{schaffner2024community}, which extracted the five sentences before and after the sentence containing keywords (See Table \ref{tab:keywords} for the keywords we used here) and merged the overlapping passages. Then, we filtered the collected data to retain only unique web pages and passages related to AI-generated content within the social media community. For instance, we excluded pages describing AI and generative AI used for content moderation and recommendation systems. 
All pages and passages were checked by two researchers independently for relevance, and we retained 431 passages from 361 pages of 29 platforms, forming the final dataset for analysis. We note that the remaining 11 platforms do not have pages in our dataset, indicating that they do not explicitly state any information about AI-generated content in their community. Also, two platforms had platform features related to AI-generated content, but did not explicitly describe the governance of AI-generated content for users.

Our data analysis had two steps. First, we annotated our dataset to calculate descriptive statistics of the web page components we collected. Then, we conducted qualitative analysis on our dataset, where we referred to existing content moderation policy frameworks from prior works to construct the codebook \cite{schaffner2024community, singhal2023sok}.

\subsubsection{Dataset Annotation} 
\label{sec:annotation}
To characterize the composition of our dataset, we annotated each page and its corresponding passages. Specifically, this annotation process helped to summarize what and how social media platforms explicitly respond to generative AI and AI-generated content. Based on our manual exploration before data collection, a review of the collected data set, and research team discussions, we annotated pages and passages with two main codes, respectively: \textsc{Types of Pages} and \textsc{Types of AI-Generated Content} in addition to subcodes as shown in Table \ref{tab:annotation}.

All labels were non-mutually exclusive. For example, if a policy page is contained in the support center, the page would be annotated using both \textsc{Legal and Policy} and \textsc{Support and Help Center} labels. Two researchers annotated all pages and passages in the dataset independently, and all conflicts were resolved through team discussions. Since all discrepancies were addressed during the annotation process, we did not calculate inter-rater reliability (IRR)~\cite{10.1145/3359174, armstrong1997place}.

\begin{table*}[htbp]
\caption{Dataset annotation labels}
\label{tab:annotation}
\renewcommand{\arraystretch}{1.25}
\begin{tabular}{p{0.32\textwidth}|p{0.55\textwidth}}
\toprule
\textbf{Label}                                                            & \textbf{Description}                 \\
\midrule
\textsc{Types of Pages}                                                                         
\\
Legal and Policy &  Formal legal and policy documents. e.g., Terms of Service, Community Guidelines, and Generative AI Policy
\\
Support and Help Center &  Platform support pages. e.g., Support Center and Transparency Center 
\\
News and Announcements & Platform informal documents. e.g., articles from Newsroom and Official Blog 
\\ 
Other & Pages that could not be categorized with any label above
\\
\midrule
\textsc{Types of AI-Generated Content*}  &  
\\
User-Posted AI-Generated Content & Governance on users creating and using AI-generated content in general
\\
Integrated AI Tool & Platforms provide user-facing generative AI tools for content creation
\\
Platform-Posted AI-Generated Content & Platforms produce their own AI-generated content for the community \\
\bottomrule
\multicolumn{2}{l}{\footnotesize *Indicates the type of AI-generated content described in the passage.}
\end{tabular}
\end{table*}

\begin{table*}
\centering
\caption{Truncated codebook reflecting the high-level structure, with sub-codes developed from content moderation frameworks for the parent codes with an *.}
\label{tab:codebook}
\renewcommand{\arraystretch}{1.25}
\begin{tabular}{p{0.32\textwidth}|p{0.55\textwidth}}
\toprule
\textbf{Parent Code} & \textbf{Description} 
\\
\midrule
\textsc{AI-Generated Content Governance*} & Actions related to AI-generated content users create and use in general
\\
\textsc{Integrated AI Tool Governance*} & Actions related to AI-generated content platforms make
\\
\textsc{User Empowerment} & Platforms provide materials for users about AI-generated content
\\
\midrule
\textbf{Sub-Code (apply to parent codes with *)}                                                            & \textbf{Description}                       
\\
\midrule
\textsc{Governance Rationale} & Why platforms govern AI-generated content
\\
\textsc{User Rules} & Rules on creating, posting, and sharing AI-generated content
\\
\textsc{Detection Methods} & How platforms identify AI-generated content and rule violations
\\
\textsc{Governance Consequence} & Enforcement on AI-generated content rules and user appeals
\\
\bottomrule
\end{tabular}
\end{table*}

\subsubsection{Qualitative Thematic Analysis}
\label{subsec:qualanalysis}
We conducted thematic analysis on the passages in our dataset through two rounds of coding \cite{thematicanalysis, braun2021thematic} using the qualitative analysis tool MAXQDA.\footnote{\url{https://www.maxqda.com/}} The resulting codes were applied to 27/29 platforms. First, since we were interested in governance around AI-generated content rules created and posted by users, we focused only on passages in the annotated dataset labeled as ``User-posted AI-Generated Content'' and ``Integrated AI Tool''. We began by familiarizing ourselves with the annotated data to develop the initial codebook with two main parent codes: \textsc{AI-Generated Content Governance} and \textsc{Integrated AI Tool Governance}. We also referred to two frameworks from prior work that describe and categorize online platforms' typical content moderation rules to help us understand what platform governance rules may cover while reading through the data. The first framework by Schaffner et al. on platforms' content moderation policies for user-generated content~\cite{schaffner2024community} covers five key elements: policy justifications on why platforms moderate; moderation criteria on what platforms moderate; safeguards on how violations are detected; platform responses on the outcomes of being moderated; and appeal/redress on how users can contest moderation decisions. The second framework by Singhal et al. for social media platforms content moderation rules covers three elements ~\cite{singhal2023sok}: moderation policies to define what needs to be moderated; detection methods to find policy violations; and enforcement of moderation policies for content and users.

One researcher then used inductive coding on passages from a partial set of the data (nine platforms or approximately 30\% of the total passages that needed to be analyzed) and further refined the initial codebook after team discussions. During this process, we added another parent code, \textsc{User Empowerment}, to cover any platform actions related to user education and controls over AI-generated content, which prior work considers to be part of a governance response~\cite{obrien2025dating}. We also added four sub-codes based on content moderation frameworks (see Table \ref{tab:codebook} for an overview of the codebook). Using this codebook, two researchers then coded all passages in the dataset, with one serving as the primary coder and one serving as the secondary coder. During this second round of coding on the entire dataset, we also created additional sub-codes to the top-level sub-codes to further differentiate new concepts (see Table \ref{tab:finalcode1}, Table \ref{tab:finalcode2}, and Table \ref{tab:finalcode3} in the Appendix for the complete final codebook). We resolved discrepancies during coding through regular meetings and team discussions to achieve consensus about points of disagreement. Since we used the coded passages as input to our thematic analysis, we did not calculate IRR~\cite{10.1145/3359174, armstrong1997place}. 

After the data was coded, we developed thematic summaries for the main codes of interest using data extracts to highlight themes. We then discussed these summaries with the research team to derive the paper themes around six main governance strategies on AI-generated content.

\subsection{Limitations and Ethics}

First, our data was collected by large-scale web scraping of publicly available web pages. We may have missed relevant documents that are not accessible to web scraping (e.g., PDF files), excluded pages due to dynamic web structures or blocking, and omitted pages that have since been modified or updated. Second, the data represent a one-time snapshot at the time of collection, and thus cannot capture long-term changes in AI governance approaches in social media platforms. Meanwhile, our data collection only covered the 40 most popular social media platforms, so smaller platforms may have different approaches than what we observed. We also acknowledge that our method can only capture platforms that explicitly state their governance actions with AI-related terms around, and platforms' exact actions can be inconsistent with what they claim.

Since our study does not involve human subjects, we did not require an Institutional Review Board (IRB) approval from our institution. We note that when performing scraping, we followed the best ethical practices for scraping for research purposes, such as respecting robots.txt. Specifically, we tweaked our scraper to prevent any scraping of pages containing user profiles or identifiable information.

\section{Dataset Overview: Social Media Features and Governance of AI-Generated Content}
\label{sec:data}

Our dataset reveals that social media platforms are proactively introducing governance of AI-generated content to mitigate misuse and encourage responsible creation and engagement. Table \ref{tab:metastats1} in the Appendix presents an overview of the presence of generative AI governance and features in the pages and corresponding passages we collected, across the 40 social media platforms we studied. 

Across social media platforms we studied, we identified 27/40 platforms that explicitly govern, i.e., describe their rules and enforcement in the collected pages on how users create, share, and discuss AI-generated content. The other 13/40 platforms do not have any explicit descriptions of how they govern AI-generated content specifically, meaning this content is subject to existing policies implicitly.

Most of the information on the governance of posted AI-generated content is scattered in Legal and Policy pages, Support pages such as Help Center and Transparency Center, and on News, Announcements, or Official Blogs (see Tables \ref{tab:location} in Appendix for detailed summary). Only 4/27 platforms (Medium, Patreon, StackOverflow, TikTok) have established a formal policy page stating key points on governance of AI-generated content---summarizing essential information on user rules and enforcement in all of their governance actions.

17/40 platforms provide their own integrated AI tools for social media users, including general-purpose Large Language Model (LLM) chatbots or profile-building assistants (6/17 platforms); tools for creators to generate creative content (14/17 platforms); and tools for business services, e.g., to help create AI-generated advertisements (7/17 platforms). Most tools offer the ability to generate text, image, video, and audio, and can be used for full AI generation, as an AI assistant, or to make minor edits to content with AI effects. For example, Facebook and other Meta platforms integrate a standalone Meta AI chatbot portal into their platforms,\footnote{\url{https://www.facebook.com/help/994963392422769}} provide features such as AI-based writing assistants and photo expansions for users to create posts,\footnote{\url{https://www.facebook.com/help/7745061278859290}} and incorporate tools for making creative AI-generated content in advertiser marketing tools.\footnote{\url{https://www.facebook.com/business/help/539137881899016}} At least 12/17 platforms with integrated AI tools have a page with dedicated policy terms for the use of these tools, where the governance actions on tool output generation and distribution are usually presented.  

\begin{table*}[htbp]
\centering
\caption{Summary of six governance approaches over AI-generated content we found across 27 social media platforms stating such information. The first four approaches apply to any AI-generated content users create, post, and share on the platform. The next two apply to the platform's integrated AI tools and to tools and resources to help users understand and interact with AI-generated content, respectively.}
\label{tab:govsum}
\renewcommand{\arraystretch}{1.45}
\begin{tabular}{p{0.35\textwidth}|p{0.53\textwidth}|c}
\toprule
\textbf{Governance Approach}                                                                              & \textbf{Description}                                                                                      &      \textbf{Count}                                                                                                   \\
\midrule
Moderating AI-Generated Content That Violates Existing Policies \newline(Existing Policies, Section \ref{subsec: 5.1})              & Platforms deal with inappropriate AI-generated content using their existing content moderation policies included in Community Guidelines and Terms of Service.& 25                         \\

Disclosing and Labeling AI-Generated Content \newline(Disclosing/Labeling, Section \ref{subsec: 5.2})                              & Platforms require or encourage users to disclose AI-generated content. Platforms also apply labels to AI-generated content automatically when disclosure is not required.& 18                                                    \\
Restricting Posting and Sharing AI-Generated \newline Content \newline(Specific Restrictions, Section \ref{subsec: 5.3})                       & Platforms impose additional restrictions around the ownership, value, and quality of AI-generated content.& 5 \\
Constraining Monetization of AI-Generated \newline Content \newline(Monetization, Section \ref{subsec: 5.4})                          & Platforms constrain the monetization of AI-generated content based on general policies around posting and sharing such content.&  6                                                                                       \\
Controlling Output Generation and Distribution \newline from Integrated AI Tools \newline(Tool Output, Section \ref{subsec: 5.5})    & Platforms moderate their integrated AI tools' output generation and regulate sharing of this content to the community.& 14                                                                                 \\
Educating and Empowering Users When Interacting with AI-Generated Content \newline(User Empowerment, Section \ref{subsec: 5.6})  & Platforms provide users with tools and resources to enhance user autonomy to distinguish and control AI-generated content in their feeds, and develop critical literacy to responsibly post and share such content.& 17\\
\bottomrule
\end{tabular}
\end{table*}
\begin{table*}[t]
\centering
\caption{Presence of platform governance approaches over AI-generated content, across 27 social media platforms stating such information. }
\label{tab:userpostai}
\resizebox{\textwidth}{!}{
\renewcommand{\arraystretch}{1.15}
\setlength{\tabcolsep}{3pt}
\begin{tabular}{p{80pt}*{27}{p{13pt}}}
\toprule
\rule{0pt}{1.8cm}&
\rothead{Behance} & \rothead{Dailymotion} & \rothead{DeviantArt} & \rothead{Facebook} &
\rothead{Flickr} & \rothead{Instagram} & \rothead{LinkedIn} & \rothead{Medium} &
\rothead{OnlyFans} & \rothead{Patreon} & \rothead{Pinterest} & \rothead{Pixabay} &
\rothead{Pixiv} & \rothead{Pornhub} & \rothead{Quora} & \rothead{Reddit} & \rothead{Roblox}& \rothead{SoundCloud} &
\rothead{StackOverflow} & \rothead{Threads} & \rothead{TikTok} & \rothead{Twitch} &
\rothead{Twitter/X} & \rothead{Vimeo} & \rothead{XHamster} & \rothead{Yelp} & \rothead{YouTube} \\
\midrule
Existing Policies
  & \textcolor{ForestGreen}{\ding{51}} 
  & \textcolor{ForestGreen}{\ding{51}} 
  & \textcolor{ForestGreen}{\ding{51}} 
  & \textcolor{ForestGreen}{\ding{51}} 
  & \textcolor{ForestGreen}{\ding{51}} 
  & \textcolor{ForestGreen}{\ding{51}} 
  & \textcolor{ForestGreen}{\ding{51}} 
  & \textcolor{ForestGreen}{\ding{51}} 
  & \textcolor{ForestGreen}{\ding{51}} 
  & \textcolor{ForestGreen}{\ding{51}} 
  & \textcolor{ForestGreen}{\ding{51}} 
  & \textcolor{ForestGreen}{\ding{51}} 
  & \textcolor{ForestGreen}{\ding{51}} 
  & \textcolor{ForestGreen}{\ding{51}} 
  & \textcolor{gray!70}{\ding{55}}
  & \textcolor{ForestGreen}{\ding{51}} 
  & \textcolor{gray!70}{\ding{55}}
  & \textcolor{ForestGreen}{\ding{51}} 
  & \textcolor{ForestGreen}{\ding{51}}   
  & \textcolor{ForestGreen}{\ding{51}} 
  & \textcolor{ForestGreen}{\ding{51}} 
  & \textcolor{ForestGreen}{\ding{51}} 
  & \textcolor{ForestGreen}{\ding{51}} 
  & \textcolor{ForestGreen}{\ding{51}} 
  & \textcolor{ForestGreen}{\ding{51}} 
  & \textcolor{ForestGreen}{\ding{51}}    
  & \textcolor{ForestGreen}{\ding{51}} 
\\
Disclosing/Labeling
  & \textcolor{ForestGreen}{\ding{51}} 
  & \textcolor{gray!70}{\ding{55}}    
  & \textcolor{ForestGreen}{\ding{51}} 
  & \textcolor{ForestGreen}{\ding{51}} 
  & \textcolor{ForestGreen}{\ding{51}} 
  & \textcolor{ForestGreen}{\ding{51}} 
  & \textcolor{ForestGreen}{\ding{51}} 
  & \textcolor{ForestGreen}{\ding{51}} 
  & \textcolor{ForestGreen}{\ding{51}} 
  & \textcolor{gray!70}{\ding{55}}    
  & \textcolor{ForestGreen}{\ding{51}} 
  & \textcolor{ForestGreen}{\ding{51}} 
  & \textcolor{ForestGreen}{\ding{51}} 
  & \textcolor{gray!70}{\ding{55}}    
  & \textcolor{gray!70}{\ding{55}}
  & \textcolor{ForestGreen}{\ding{51}} 
  & \textcolor{gray!70}{\ding{55}}
  & \textcolor{gray!70}{\ding{55}} 
  & \textcolor{gray!70}{\ding{55}} 
  & \textcolor{ForestGreen}{\ding{51}} 
  & \textcolor{ForestGreen}{\ding{51}} 
  & \textcolor{gray!70}{\ding{55}}    
  & \textcolor{ForestGreen}{\ding{51}} 
  & \textcolor{ForestGreen}{\ding{51}} 
  & \textcolor{ForestGreen}{\ding{51}} 
  & \textcolor{gray!70}{\ding{55}} 
  & \textcolor{ForestGreen}{\ding{51}} 
\\
Specific Restrictions
  & \textcolor{gray!70}{\ding{55}} 
  & \textcolor{gray!70}{\ding{55}}    
  & \textcolor{gray!70}{\ding{55}} 
  & \textcolor{gray!70}{\ding{55}} 
  & \textcolor{gray!70}{\ding{55}} 
  & \textcolor{gray!70}{\ding{55}} 
  & \textcolor{gray!70}{\ding{55}} 
  & \textcolor{ForestGreen}{\ding{51}} 
  & \textcolor{ForestGreen}{\ding{51}} 
  & \textcolor{gray!70}{\ding{55}}    
  & \textcolor{gray!70}{\ding{55}} 
  & \textcolor{ForestGreen}{\ding{51}} 
  & \textcolor{gray!70}{\ding{55}} 
  & \textcolor{gray!70}{\ding{55}}    
  & \textcolor{gray!70}{\ding{55}}
  & \textcolor{gray!70}{\ding{55}} 
  & \textcolor{gray!70}{\ding{55}}
  & \textcolor{ForestGreen}{\ding{51}} 
  & \textcolor{ForestGreen}{\ding{51}} 
  & \textcolor{gray!70}{\ding{55}} 
  & \textcolor{gray!70}{\ding{55}} 
  & \textcolor{gray!70}{\ding{55}}    
  & \textcolor{gray!70}{\ding{55}} 
  & \textcolor{gray!70}{\ding{55}} 
  & \textcolor{gray!70}{\ding{55}} 
  & \textcolor{ForestGreen}{\ding{51}} 
  & \textcolor{gray!70}{\ding{55}} 
\\
Monetization
  & \textcolor{gray!70}{\ding{55}} 
  & \textcolor{gray!70}{\ding{55}} 
  & \textcolor{ForestGreen}{\ding{51}} 
  & \textcolor{gray!70}{\ding{55}} 
  & \textcolor{gray!70}{\ding{55}} 
  & \textcolor{gray!70}{\ding{55}} 
  & \textcolor{gray!70}{\ding{55}} 
  & \textcolor{ForestGreen}{\ding{51}} 
  & \textcolor{gray!70}{\ding{55}} 
  & \textcolor{ForestGreen}{\ding{51}} 
  & \textcolor{gray!70}{\ding{55}} 
  & \textcolor{gray!70}{\ding{55}} 
  & \textcolor{ForestGreen}{\ding{51}} 
  & \textcolor{gray!70}{\ding{55}} 
  & \textcolor{gray!70}{\ding{55}}
  & \textcolor{ForestGreen}{\ding{51}} 
  & \textcolor{gray!70}{\ding{55}}
  & \textcolor{ForestGreen}{\ding{51}} 
  & \textcolor{gray!70}{\ding{55}} 
  & \textcolor{gray!70}{\ding{55}} 
  & \textcolor{gray!70}{\ding{55}} 
  & \textcolor{gray!70}{\ding{55}}    
  & \textcolor{gray!70}{\ding{55}} 
  & \textcolor{gray!70}{\ding{55}} 
  & \textcolor{gray!70}{\ding{55}} 
  & \textcolor{gray!70}{\ding{55}} 
  & \textcolor{gray!70}{\ding{55}} 
\\
Tool Output
  & \textcolor{ForestGreen}{\ding{51}} 
  & --- 
  & \textcolor{ForestGreen}{\ding{51}} 
  & \textcolor{ForestGreen}{\ding{51}} 
  & --- 
  & \textcolor{ForestGreen}{\ding{51}} 
  & \textcolor{ForestGreen}{\ding{51}} 
  & --- 
  & --- 
  & --- 
  & \textcolor{ForestGreen}{\ding{51}} 
  & --- 
  & --- 
  & --- 
  & \textcolor{ForestGreen}{\ding{51}}
  & \textcolor{gray!70}{\ding{55}} 
  & \textcolor{ForestGreen}{\ding{51}}
  & \textcolor{ForestGreen}{\ding{51}} 
  & \textcolor{gray!70}{\ding{55}} 
  & \textcolor{ForestGreen}{\ding{51}} 
  & \textcolor{ForestGreen}{\ding{51}} 
  & ---    
  & \textcolor{ForestGreen}{\ding{51}} 
  & \textcolor{ForestGreen}{\ding{51}} 
  & --- 
  & --- 
  & \textcolor{ForestGreen}{\ding{51}} 
\\
User Empowerment
  & \textcolor{gray!70}{\ding{55}} 
  & \textcolor{gray!70}{\ding{55}} 
  & \textcolor{ForestGreen}{\ding{51}} 
  & \textcolor{ForestGreen}{\ding{51}} 
  & \textcolor{gray!70}{\ding{55}} 
  & \textcolor{ForestGreen}{\ding{51}} 
  & \textcolor{ForestGreen}{\ding{51}} 
  & \textcolor{ForestGreen}{\ding{51}} 
  & \textcolor{gray!70}{\ding{55}} 
  & \textcolor{gray!70}{\ding{55}} 
  & \textcolor{ForestGreen}{\ding{51}} 
  & \textcolor{ForestGreen}{\ding{51}} 
  & \textcolor{ForestGreen}{\ding{51}} 
  & \textcolor{ForestGreen}{\ding{51}} 
  & \textcolor{gray!70}{\ding{55}} 
  & \textcolor{ForestGreen}{\ding{51}} 
  & \textcolor{gray!70}{\ding{55}} 
  & \textcolor{gray!70}{\ding{55}} 
  & \textcolor{ForestGreen}{\ding{51}} 
  & \textcolor{ForestGreen}{\ding{51}} 
  & \textcolor{ForestGreen}{\ding{51}} 
  & \textcolor{ForestGreen}{\ding{51}}    
  & \textcolor{ForestGreen}{\ding{51}} 
  & \textcolor{ForestGreen}{\ding{51}} 
  & \textcolor{gray!70}{\ding{55}} 
  & \textcolor{gray!70}{\ding{55}} 
  & \textcolor{ForestGreen}{\ding{51}} 
\\
\bottomrule
\multicolumn{28}{p{1.15\textwidth}}{\textcolor{ForestGreen}{\ding{51}}: passage(s) about this topic exists in our dataset, \textcolor{gray!70}{\ding{55}}: passage(s) about this topic does not exist in our dataset, ---: Not applicable for `Tool Output', since the platform does not have an integrated AI tool. }
\end{tabular}}
\end{table*}

Interestingly, 10/40 platforms we studied use generative AI to summarize user-generated content, such as articles and videos produced by creators; to search and summarize user-generated content over the community based on users' queries; produce AI-generated content for the platform; or use AI to aid with accessibility such as converting text to AI-generated audio. For example, YouTube provides AI-generated video summaries and offers an AI chatbot that allows users to ask questions about the video content.\footnote{\url{https://blog.youtube/inside-youtube/2024-in-youtube-ai}} In another example, Quora deploys AI-generated answers for selected questions posted by users.\footnote{\url{https://help.quora.com/hc/en-us/articles/12677380466068-What-are-Sage-and-Dragonfly-bots-on-Quora}}

\section{Findings: Six Governance Approaches Over AI-Generated Content in Social Media}
\label{sec:findings} 

We identified six main governance approaches from the 27 social media platforms that describe their governance of AI-generated content, as shown in Table \ref{tab:govsum} and Table~\ref{tab:userpostai}. Four of the governance approaches cover any AI-generated content posted and shared by users; one covers the use of integrated AI tools on the platform; and one covers how platforms empower users to interact with AI-generated content and integrated AI tools.

Of the six approaches, the most common ones used in social media platforms are moderating AI-generated content for appropriateness and authenticity---as they would any other content according to existing content moderation policies (Section \ref{subsec: 5.1})---and disclosing and labeling AI-generated content (Section \ref{subsec: 5.2}). A smaller number of platforms go further in their governance actions: imposing AI-specific posting and sharing restrictions when content is not otherwise inappropriate---i.e., violating any existing content moderation policy (Section \ref{subsec: 5.3})---and constraining monetization of AI-generated content (Section \ref{subsec: 5.4}). For platforms with integrated AI tools, we found that many apply specific safety measures to the tool output generation and distribution (Section \ref{subsec: 5.5}). Lastly, some platforms provide features and resources to help users recognize and critically evaluate AI-generated content, as well as guide them in creating and sharing this content responsibly (Section \ref{subsec: 5.6}). 

\paragraph{Definitions of AI-Generated Content by Social Media Platforms} Before detailing the six governance approaches we observed, we first briefly summarize how platforms define AI-generated content. Across platforms we studied, terminology such as AI-generated, AI-assisted, deepfake, or synthetic media often appears in policies and support documentation, but without clear or consistent definitions. We found only nine platforms (DeviantArt, Medium, Patreon, Pixiv, Pornhub, StackOverflow, TikTok, Twitter/X, YouTube) that explicitly define AI-generated content and related concepts. These definitions, however, are varied across platforms. For example, TikTok defines AI-generated content as broadly any content created with any degree of AI involvement: \textit{``AI-generated content includes images, video, and/or audio that is generated or modified by deep- or machine-learning processes.''} (TikTok’s Support Center article). In contrast, some platforms define AI-generated content only as content with significant AI components: \textit{``We define `AI-generated' work as work in which all or most of the production process is carried out by AI.''} (Pixiv's Help Center article). 

Beyond general AI-generated content, a few platforms also define deepfake and synthetic media, while usually emphasizing it as content that has been manipulated for a particular purpose rather than the fact that the content is AI-generated: \textit{``Deepfakes are manipulated or digitally-altered audio, video, or image works that portray a person, event, or scene in a way that is not real or did not occur.''} (Patreon's Community Guidelines).  

In short, many platforms do not have an explicit definition of AI-generated content and related terms. These definitions, whenever they do exist, are usually varied across platforms.   

\subsection{Moderating AI-Generated Content That Violates Existing Policies} 
\label{subsec: 5.1}
We found 25 platforms explicitly outline their governance approach to inappropriate AI-generated content that violates existing, often well-developed content moderation policies (i.e., Terms of Service and Community Guidelines). Their goals in this case are the same as for any content that needs to be moderated for compliance with laws and to keep their community safe. To achieve these goals, platforms apply existing moderation criteria regarding the prohibition of AI-generated content, use both platform-driven and user-driven methods to detect violations, and take strict actions against both offending AI-generated content and users, while offering some degree of appeals for moderated users. 

Below, we present user rules on AI-generated content violating existing policies in the platform, detection methods for such violations, and moderation consequences. We note that, even though the other 15 of the 40 platforms we studied do not explicitly mention AI-generated content, their existing content moderation policies are still applicable to this content~\cite{fisher2024moderating}. Our analysis only focuses on whether and how AI-generated content is highlighted in a platform's existing content moderation policies. 

\subsubsection{User Rules on AI-Generated Content Violating Existing Policies} Aligning with their community guideline restrictions, all 25 social media platforms highlight their strict prohibition of posting AI-generated content that could be considered illegal or compromising community safety. 
Structurally, we found that rather than forming separate AI-specific rules, platforms have added AI-generated content as a condition to existing rules on user-generated content. For instance, \textit{``You cannot impersonate or misrepresent yourself as another person (including through artificial intelligence) or post information or content that is intentionally misleading.''} (Soundcloud's Community Guideline). 
In particular, eight platforms (Behance, Facebook, Instagram, OnlyFans, Pinterest, Threads, TikTok, YouTube) separately emphasize that AI-generated content posted by users is a part of user-generated content that must comply with existing platform terms and guidelines. As represented by the following example, \textit{``Our Community Standards apply to everyone, all around the world, and to all types of content, including AI-generated content.''} (Meta's Community Standards applied to Facebook, Instagram, and Threads).

Prior works studying content moderation policies and community guidelines show that platforms categorize various types of prohibited content that is considered inappropriate at differing levels of granularity~\cite{singhal2023sok, jiang2020characterizing, arora2023detecting, schaffner2024community}. In our analysis, much like how content is usually moderated on platforms with user-generated content, social media platforms typically say that AI-generated content is prohibited if it constitutes misleading information (e.g., spam, scam, and impersonation), hateful content (e.g., bullying and humiliation), copyright infringement and plagiarism, privacy violations, and harmful illegal content (e.g., child abuse materials and non-consensual intimate media). 

Moreover, platforms forbid AI-generated content that violates existing rules regardless of its modality and the role AI plays in its creation. As emphasized in the following rule, \textit{``Any `deepfakes' whatsoever (AI-generated, modified, or other manipulations of a person's image, whether in picture or video, to deceive or mislead the viewer into believing that person is acting or speaking in the way presented) [are prohibited].''} (Pornhub's Non-Consensual Content Policy). We also noticed three platforms (Medium, Reddit, Patreon) stress the prohibition of redirection to AI-generated content and generative AI services that violate existing rules but are hosted outside the platform, \textit{``A new generation of apps leveraging AI to generate nonconsensual nude images of real people have emerged across the Internet. To be very clear: sharing links to these apps or content generated by them is prohibited on Reddit.''} (Reddit's Announcement on r/modnews).

\subsubsection{Detection of Policy-Violating AI-Generated Content.} 
Social media platforms in our dataset primarily state the same detection strategies for typical content that violates existing policies to address inappropriate AI-generated content. Platforms typically use a combination of automatic detection and human review to find inappropriate content, and largely rely on user reports to supplement their own efforts~\cite{schaffner2024community, singhal2023sok}. As such, how platforms detect policy-violating AI-generated content is rarely emphasized separately. 

We found that five platforms (Facebook, Instagram, Reddit, Thread, YouTube) also detail their detection techniques for AI-generated content violating existing policies. For example, \textit{``We also have a network of nearly 100 independent fact-checkers who will continue to review false and misleading AI-generated content.''} (Meta's Newsroom article applies to Facebook, Instagram, and Threads). We also identified new detection tools developed to address specific violations of AI-generated content, going beyond general moderation methods. One example is a likeness matching technique for detecting deepfakes infringing privacy and copyright: \textit{``we've developed likeness management technology, including new synthetic-singing identification technology within Content ID that will allow partners to automatically detect and manage AI-generated content on YouTube that simulates their singing voices, and new technology that will enable people from a variety of industries [...] to detect and manage AI-generated content showing their faces on YouTube.''} (YouTube's Official Blog).

Additionally, three platforms (DeviantArt, TikTok, YouTube) explicitly mention the role of users to help identify and report inappropriate AI-generated content, with one emphasizing that the reporting process mirrors that for any other policy-violating user-generated content: \textit{``As with any deviation, if you come across inappropriate AI-generated content, please use the Report option in the `...' menu.''} (Deviantart's Official Blog).

\subsubsection{Consequences of Violations on Existing Policies.} When users post AI-generated content that violates community guidelines, social media enforcement actions are typically the same as for general content moderation enforcement~\cite{schaffner2024community, singhal2023sok}. These consequences for AI-generated content, highlighted by 11 platforms, include hard moderation, like removing the offending content or restricting the account that posted it, and soft moderation, such as flagging content that intends to mislead people. Similar to how AI-generated content is framed as a form of user-generated content subject to existing policies, we found that four platforms (Facebook, Instagram, Threads, TikTok) emphasize that their enforcement against inappropriate AI-generated content and its creator is the same as for any rule-violating content. As exemplified below, \textit{``We take action on any content that violates our Community Guidelines, regardless of whether it was altered with AI. This includes our policies around impersonation, misinformation, and hate speech.''} (TikTok's Support Center article). 

With regard to user appeal and redress mechanisms for AI-generated content that has been moderated, platforms in our dataset also generally offer the same processes as those used for general policy violations. However, we only observed two platforms (Reddit, Twitter/X) describing appeal and redress approaches when talking about the moderation of AI-generated content, with neither platform explicitly explaining the handling of AI-generated content. This lack of information on user appeals and redress for AI-generated content being moderated is unsurprising, given the limited coverage of appeal and redress mechanisms in content moderation policies in general~\cite{schaffner2024community}. 

\subsubsection{Takeaways}
Many platforms emphasize AI-generated content as a subset of user-generated content, subject to existing user rules, detection methods, and enforcement mechanisms. This approach positions AI-generated content not as a fundamentally new governance objective, but as an emerging focus of the existing content moderation pipeline. At the policy level, this framing locates accountability in the actions of users who create and distribute AI-generated content, rather than in the AI systems themselves or the AI generation process.

\subsection{Disclosing and Labeling AI-Generated Content} 
\label{subsec: 5.2}
In our dataset, 18 social media platforms adopt governance principles on disclosing and labeling of AI-generated content. This governance approach responds to the growing use of generative AI tools for content creation, aiming to safeguard content authenticity---even where creators have no malicious intent---and clarifying distinctions in authorship between human-made and AI-generated content. Specifically, platforms require or encourage users to disclose AI-generated content they post, enforce these rules by identifying if the content is AI-made or not, and adopt looser enforcement for violations of disclosure rules compared to rules on content violating existing policies. Even where disclosing AI-generated content is not required, platforms also state that they may enforce measures such as auto-labeling this content. Notably, four platforms (LinkedIn, Pinterest, Twitter/X, XHamster) impose no user disclosure rule, yet still describe their detection and auto-labeling practices on AI-generated content. Below, we describe user rules on disclosing AI-generated content, detection of AI-generated content, and governance consequences related to labeling AI-generated content.

\subsubsection{User Rules on Disclosing AI-Generated Content} 
We found 14 platforms in our dataset that have established rules around disclosing AI involvement when users post and share content. Most platforms claim that these disclosure rules are intended to further promote content authenticity, transparency, and trust within their communities. As stated in Instagram's Help Center: \textit{``Labeling your AI-generated or AI-modified content on Instagram helps establish transparency and trust. In some cases it's required.''} (Instagram's Help Center article). We found two platforms (OnlyFans, Pixabay) that require any AI-generated content to be disclosed without exception. The following rule exemplifies this: \textit{``AI Generated content must comply with our Terms of Service and must be clearly and conspicuously captioned as AI Generated Content with a signifier such as \#ai, or \#AIGenerated.''} (OnlyFans' Terms of Service). 

For other platforms, however, the disclosure of AI-generated content is conditional or not mandatory. In six platforms (Facebook, Instagram, YouTube, Threads, Vimeo, TikTok), the emphasis on disclosure is driven by the increasing capability of generative AI models to create realistic media that may mislead users. 
As such, these platforms typically require disclosure of AI-generated content that appears to be realistic, while excluding content that is unrealistic or that has minor edits with AI from the requirements. As described below, \textit{``[We require] creators to disclose to viewers when realistic content – content a viewer could easily mistake for a real person, place, scene, or event – is made with altered or synthetic media, including generative AI. We're not requiring creators to disclose content that is clearly unrealistic, animated, includes special effects, or has used generative AI for production assistance.''} (YouTube's Official Blog). This obligatory disclosure is emphasized when the content covers sensitive topics: \textit{``We're also updating this policy to require that political ads containing AI-generated content be clearly labeled as such.''} (Reddit's Official Blog). Moreover, since concerns around realistic AI-generated content primarily relate to image, audio, or video, platforms often state that these disclosure requirements here are typically applied to multi-modal content, or do not specify the medium of AI-generated content that has to be disclosed in their policies.


Whereas disclosing realistic AI media seeks to protect content authenticity, three creativity platforms (DeviantArt, Medium, Pixiv) establish disclosure requirements focusing on content authorship. They aim to separate machine-made content from human creations, enabling users to recognize the different origins of creative works and protect human creators' rights. As such, these platforms require disclosure where AI is significantly involved in the creative process, extending across various content formats. As exemplified in the disclosure requirements: \textit{``The Service defines `AI-generated' work as work in which all or most of the production process is carried out by AI (artificial intelligence). All AI-generated work posted to the Service must be labeled as such. [...] For example, work created by combining specific prompts and posted with little to no editing, regardless of format (e.g., text, images).''} (Pixiv's Help Center article). The criteria for disclosing AI-generated content here depend on each platform's subjective definition of significant AI involvement, resulting in variation across platforms' standards. For example, Deviantart states that the disclosure of AI-generated content is not needed \textit{``if the main focus of the image is self-created and it only includes a small, insignificant AI-generated component''}; while Medium requires the disclosure of any AI-generated text and image, with only the exception of \textit{``the use of AI-assistance such as grammar or spell checkers, outline assistance, or fact verification''}. 

Additionally, we found two platforms (Behance, Flickr) solely \textit{``encourage''} or \textit{``recommend''} users to disclose any content created with AI, rather than establishing clear disclosure boundaries or enforcing strict disclosure requirements. 


\subsubsection{Distinguishing AI-Generated Content} 
The enforcement of the user disclosure rule, as well as other governance actions around labeling AI-generated content, largely relies on the ability to distinguish AI-generated content from other content. 13 social media platforms with rules on disclosing and labeling AI-generated content also describe their development and adoption of techniques to identify such content on their own. 

Mostly, platforms adopt automatic techniques to distinguish AI-generated content, or make claims about their efforts in developing such mechanisms. For example, \textit{``Our long-term goal is to develop automated labeling systems that can reliably detect AI-generated content, further enhancing transparency and reducing the burden on creators.''} (Vimeo's Official Blog). We identified two types of detection in platforms providing technical details. The first approach, mentioned by nine platforms (Behance, DeviantArt, Facebook, Instagram, LinkedIn, Pinterest, Threads, TikTok, YouTube), relies on finding digital signatures attached to the multi-modal AI-generated content during its production, such as watermarks in metadata and content credentials. These digital signatures are often attached by mainstream generative AI services, not only revealing whether the content is AI-generated but also providing additional information about its origin and production. As elaborated by Meta's statement on how the detection of invisible signals works: \textit{``We’re building industry-leading tools that can identify invisible markers at scale – specifically, the `AI generated' information in the C2PA and IPTC technical standards – so we can label images from Google, OpenAI, Microsoft, Adobe, Midjourney, and Shutterstock as they implement their plans for adding metadata to images created by their tools.''} (Meta's Newsroom article applies to Facebook, Instagram, and Threads). 

The second approach is algorithmic detection through newly developed classifiers and the existing moderation systems, as described by five platforms (Facebook, Instagram, Medium, Pinterest, Threads). Platforms describe algorithmic detection as a supplement to identifying content resource signals to detect AI-generated content: \textit{``We not only analyze an image’s metadata following the IPTC Metadata Standard, but also developed classifiers that automatically detect Gen AI content, even if the content doesn’t have obvious markers.''} (Pinterest's Help Center article).

In contrast to many platforms that describe technical details of automatic detection, only Medium explicitly notes the incorporation of human reviewers to help distinguish AI-generated content: \textit{``We use a wide variety of tools and technologies to detect and identify AI-writing and other AI content, combined with human review of any positive results.''} (Medium's AI Content Policy). Meanwhile, TikTok is the only platform that mentions the role of users reporting on undisclosed AI-generated content, but presents an underdeveloped report mechanism. Specifically, this platform instructs users to report AI-generated content that violates disclosure rules under \textit{``Deepfakes, synthetic media, and manipulated media''}, which is the same category for reporting general misinformation and harmful deepfakes.

\subsubsection{Consequences Related to Labeling AI-Generated Content} 
We found that social media platforms enforce the disclosure of AI-generated content by auto-labeling it, with support from detection techniques. For AI-generated content and users that violate disclosure rules, only a few platforms take measures beyond auto-labeling. This represents looser enforcement compared to the stricter content moderation applied to inappropriate AI-generated content as defined by existing content moderation rules.

In our dataset, 13 platforms state that they auto-label content that is AI-made where users fail to comply with disclosure rules, or even where user disclosure is not required. For example, \textit{``Meta does not require you to label images that have been created or altered with AI. Images will still receive a label if Meta’s systems detect they were AI-generated.''} (Facebook and Instagram's Help Center article). However, we found that such auto-labeling processes typically offer no opportunity for users to appeal or provide feedback. We only found two platforms (Pinterest, Vimeo) that provide appeal mechanisms for human-made content that is mislabeled as AI-generated. As represented below: \textit{``However, we know these systems aren’t perfect. This is why we set up an appeals system for creators and users who believe that their content may be mislabeled.''} (Pinterest's Newsroom article). Additionally, three platforms (TikTok, Vimeo, YouTube) specifically emphasize that users have no ability to remove or override AI labels manually, with only one of them providing appeal mechanisms. For example, \textit{``Once your content is labeled as AI-generated with an auto label, you won't be able to remove the label from your post.''} (TikTok's Support Center article).

When users violate disclosure rules, many platforms do not impose additional penalties beyond auto-labeling, except for five platforms (Facebook, Instagram, Vimeo, Threads, YouTube) that express a strike policy to penalize repeated non-disclosures of AI-generated content: \textit{``Creators who consistently choose not to disclose this information may be subject to penalties from Vimeo, including removal of content or account termination.''} (Vimeo's Help Center article). We also found two platforms (Vimeo, YouTube) emphasize prioritizing enforcement against AI-generated content that violates existing policies, even when such content complies with disclosure rules: \textit{``However, labeling alone may not be enough to mitigate the risk of harm. Some synthetic media, regardless of labeling, will be removed if it violates our Community Guidelines.''} (YouTube's Official Blog).

\subsubsection{Takeaways}
Platforms establish user rules on disclosing AI-generated content with varying standards, while also detecting and labeling any AI-generated content by themselves whenever possible. This governance approach treats AI-generated content as acceptable (as long as it does not violate any existing rules) while requiring additional safeguards on content transparency and authenticity. By enforcing conditional user-side disclosures and applying unconditional platform-side auto-labeling for AI-generated content, platforms position transparency around this content as a shared but asymmetrically enforced governance responsibility.

\subsection{Restricting Posting and Sharing AI-Generated Content} 
\label{subsec: 5.3}
Whether or not explicitly stated, most platforms permit users to post AI-generated content as long as it complies with existing policies. Five platforms (Medium, Pixabay, SoundCloud, StackOverflow, Yelp), however, impose additional restrictions tailored to AI-generated content extending beyond existing policies. Compared to moderation under existing policies or disclosure and labeling, such tailored restrictions are far less common and are mostly found on non-mainstream social media services focused on creative writing, art, and knowledge-sharing. This type of governance often seeks to address contentious issues rather than risks, such as ownership and user-perceived quality and value of AI-generated content. We found that platforms establish a spectrum of restrictions from requiring valid licenses for posting to completely prohibiting any AI-generated content. Platforms report a similar pattern of general moderation methods to enforce these rules and treat violations similarly to any inappropriate content that needs to be moderated, though details on detection and enforcement are often limited.

Below, we present user rules on tailored restrictions for AI-generated content, detection and mitigation of such violations, and consequences of violations.

\subsubsection{User Rules on Post Restrictions}  
We found two platforms (Pixabay, SoundCloud) highlight the user's responsibility for ensuring they have the legal right to post and use any AI-generated content: \textit{``You may upload Content you have created with generative AI technology [...] You are responsible for ensuring that the terms of the generative AI technology you have used permit you to grant the license and give the warranties outlined [above].''} (Pixabay's Terms of Service).

Moreover, we noticed more stringent, site-wide constraints on posting AI-generated content. We found two platforms (StackOverflow, Yelp) that prohibit posting any AI-generated content site-wide. To be specific, StackOverflow applies its ban of AI-generated and AI-assisted content to \textit{``all content on Stack Overflow, except each user's profile content (e.g., your "About me" text).''} Yelp, likewise, prohibits \textit{``use [of] chatbots or other AI tools to create reviews, including using such tools to draft or revise.''} Additionally, Medium, while not banning users from posting AI-generated content on the platform, excludes all fully AI-generated articles from algorithmic promotion and broader distribution.
These site-wide restrictions are often grounded in the platform’s values and positioning, whereby any AI-generated content is deemed irrelevant, of low value, or even potentially harmful to the users and community. For instance, StackOverflow cites low user demand as a key reason for prohibiting AI-generated content, alongside concerns about potential risks like hallucination: \textit{``Users who ask questions on Stack Overflow may have already sought answers elsewhere. Due to the ease of using generative artificial intelligence services, if a user wanted an answer from an artificial intelligence, they may already have sought one, and so it does not make sense to provide one here.''} (StackOverflow's Help Center article). 

\subsubsection{Detection and Enforcement}
Three platforms (Medium, SoundCloud, Yelp) describe their detection techniques on restrictive AI-specific rule violations, generally using a combination of automated methods and human review, similar to detecting any content that needs moderation~\cite{schaffner2024community, singhal2023sok}. We only found one platform, StackOverflow, where user-posted AI-generated content is prohibited site-wide, instructs users to report suspected AI-generated content to the moderator team under the category of plagiarism. Meanwhile, Medium claims there is no need for user reports when encountering AI-generated content violating restrictive rules: \textit{``If you see recommended AI-generated content in your feed or Daily Digest from an author or publication you don’t follow, use the “Show less” tool in the 3-dot menu to remove the story and retrain the algorithm. You do not need to report it to us.''} (Medium's AI Content Policy).

Similarly, the enforcement of AI-specific rule violations is similar to that of general content moderation. As explicitly mentioned by three platforms (Medium, StackOverflow, Yelp), common enforcement measures for AI-generated content violating AI-specific restrictions include content removal, shadowbanning (decreasing the prominence of this content), and account suspension. As exemplified below, \textit{``Use of chatbots or other AI tools to create reviews [...] are subject to removal and repeated violations may result in account closure.''} (Yelp's Support Center article). 

Specifically, on Medium and Yelp---where conditional or complete prohibitions apply site-wide---we observed that enforcement relies more on downranking AI-generated content in recommendations than on comprehensive detection and removal. We interpret this shift from detection to mitigation as a pragmatic response, since detecting AI-generated content, particularly text, remains unreliable and may lead to unjust enforcement~\cite{boutadjine2025human}. For example, \textit{``Our system has several strategies for minimizing the distribution of AI-generated content: some based on machine-learning, and others based on human curation, content flagging, and moderation.''} (Medium's Official Blog).

\subsubsection{Takeaways}
A small number of platforms impose AI-specific restrictions, often grounded in concerns about content and community values. This approach considers AI-generated content as misaligned with platform norms, particularly in communities emphasizing originality, expertise, or human insights. Rather than centering AI-generated content governance on harm prevention, this approach highlights preserving platform identity and human creator rights as the key governance goal.

\subsection{Constraining Monetization of AI-Generated Content}
\label{subsec: 5.4}
In our dataset, we found six platforms (DeviantArt, Medium, Patreon, Pixiv, Reddit, SoundCloud)---mostly creativity communities---that constrain the monetization of AI-generated content. The governance of monetization often addresses concerns about the quality, originality, authorship, and ownership, which are ambiguous with AI creations but critical when determining eligibility for commercial use. We identified varied constraints on monetizing AI-generated content, with many extending from rules in other governance actions. Similar to AI-specific rules, we found that the detection mechanism, platform enforcement, and user appeals targeting the violation of monetization rules are seldom mentioned and sometimes not well-developed. 

Next, we present user rules on monetizing AI-generated content, detection methods for unqualified AI-generated content for monetization, and consequences of violations.

\subsubsection{User Rules on Monetization Constraints} 
Our analysis reveals a variety of platform-imposed constraints on the monetization of AI-generated content. Some platforms extend their rules around the prohibition, restriction, and disclosure of AI-generated content to monetization, with much focusing on content ownership and authorship. For example, DeviantArt extends their disclosure requirement of content authorship, requiring any AI-generated content to be sold be labeled as such: \textit{``Where your Content is offered for sale on DeviantArt (including Commissions, Premium Downloads, Premium Galleries, Subscriptions, and Exclusives) and was created using Generative AI Programs or AI Tools, you must clearly tag or label your Content using as `Created using AI tools'.''} (DeviantArt's Terms of Service). SoundCloud, on the other hand, extends its AI-specific content rule of license requirements to monetization: \textit{``You can only distribute or monetize content to which you own the rights or have proper licensing. [...] At this time, we’re only able to approve content created with our AI integration partners for distribution or monetization.''} (Soundcloud's Help Center article). 

Three platforms (Medium, Pixiv, Reddit) explicitly prohibit monetization of any AI-generated content, citing reasons such as AI-generated content being considered low quality and having no originality. For instance, \textit{``Medium is for human storytelling, not AI-generated writing. [...] AI-generated writing (disclosed as such or not) is not allowed to be paywalled as part of our Partner Program.''} (Medium's AI Content Policy).

\subsubsection{Detection and Consequences of Violations} We found only two platforms (Medium, SoundCloud) explicitly mention their detection of AI-generated content that violates monetization constraints, where their detection mechanism is tailored to platform-specific constraints. Moreover, only Medium establishes a user report mechanism for AI-specific monetization rule violations: \textit{``If you see AI-content behind the paywall, please submit a request with a link to the story using this form .''} (Medium's AI Content Policy).

As explicitly mentioned by three platforms (DeviantArt, Medium, Reddit), common enforcement measures for AI-generated content violating monetization rules also include content-facing and user-facing enforcement, such as content removal and account restrictions. Specifically, users may have their sales refunded or have their monetization eligibility revoked: \textit{``Deviants found selling artwork that is partially or fully AI-generated without applying the `Created using AI tools' label may have their sales refunded or face disciplinary action.''} (DeviantArt's Support Center article). 

We highlight that we did not find any platforms offering appeal mechanisms for AI-generated content monetization rule violations. Medium, however, explicitly states that enforcement of monetization restrictions on AI-generated content is not subject to appeal: \textit{``Paywalling AI-generated content in violation of our AI content policy. Revocation from the Partner Program is final and there are no appeals.''} (Medium's Help Center article).

\subsubsection{Takeaways}
A few platforms restrict or prohibit monetization of AI-generated content, citing concerns around quality, originality, and ownership. This approach positions AI-generated content as economically contentious, even when its use is otherwise permitted. In doing so, these platforms differentiate between participation and compensation, allowing users to share AI-generated content while limiting its eligibility for revenue.

\subsection{Controlling Output Generation and Distribution from Integrated AI Tools}
\label{subsec: 5.5}
Our analysis shows that 14 social media platforms describe their governance actions on the output generation and distribution from their integrated AI tools. Prior work by Gao et al.~\cite{gao2025cannot} examined content moderation policies in generative AI products and found they typically cover a wide range of prohibitions on user input and AI output; detection methods on harmful input and output through automatic methods and human review; model training enhancements to refuse harmful input and reduce harmful output; and actions on both content and users who violate policies. We observed that all 14 platforms state governance actions over output generations from their integrated AI tools, focusing on tool safety and closely resembling the approaches of independent generative AI products.

However, social media platforms provide additional information on governing their integrated tools' output distribution, especially posting and sharing output within their community. Below, we present our findings on user rules and platform enforcement on the safety and responsibility of output distribution from social media platforms' integrated AI tools.

\subsubsection{User Rules on Sharing Output Within the Community}
Generative AI products typically include policies that not only restrict user inputs and system outputs, but also limit the distribution and use of outputs after they are generated. As shown in prior work~\cite{gao2025cannot}, restrictions on output use may include prohibiting a broad range of generated content usage from training competitive AI models to deceiving others into believing the content is human-made. 

In social media platforms, such rules extend to how users should publish the output into their community, as established by ten platforms (DeviantArt, Facebook, Instagram, LinkedIn, Pinterest, SoundCloud, TikTok, Vimeo, Threads, YouTube). For example, \textit{``If you use or publish any Output, you are solely responsible for evaluating the Output and determining its accuracy, suitability, completeness, and the appropriateness of using or publishing the Output on Pinterest.''} (Pinterest's Generative AI Business Terms of Service). As in the example above, users are usually instructed to double-check if the output is acceptable to share with the community, or in other words, the use of output from integrated AI tools should \textit{``adhere to the Community Guidelines''} in the same manner as for all other user-generated content. We also found extra rules beyond solely ensuring that AI-generated output is appropriate. For example, Deviantart has a disclosure requirement for art generated from its integrated AI tool and published to the platform: \textit{``If you refer to an artist in a DreamUp prompt, you must also tag that artist when submitting the resulting image to DeviantArt.''} (Deviantart's Support Center article).

\subsubsection{Governance Enforcement on Output Distribution}
To ensure the responsible use of outputs from integrated AI tools within their community, platforms also implement measures that extend beyond regulating users. The most common of these governance actions is auto-labeling such outputs when they are shared in the community, as described by nine platforms (DeviantArt, Facebook, Instagram, Pinterest, SoundCloud, Threads, TikTok, Vimeo, YouTube). For instance, \textit{``Content created by YouTube's generative AI products and features will be clearly labeled as altered or synthetic.''} (YouTube's Official Blog). Five platforms (Facebook, Instagram, Threads, TikTok, YouTube) disclose that they watermark their integrated AI tool output to enable auto-labeling within their own platform, but also to facilitate detection and labeling beyond their platform, to aid in the governance of AI content posted on other social media platforms. As stated below, \textit{``In addition, AI Alive stories will have an AI-generated label to bring transparency to how the content was created, and have C2PA metadata embedded—a kind of technology that helps others identify that it's AI-generated, even if it's downloaded and shared off platform.''} (TikTok's Newsroom article). 

Additionally, two platforms (LinkedIn, TikTok) claim an additional layer of content safety protection when users decide to share the output to the community, alongside content safety measures incorporated into the integrated AI tool. For example, \textit{``A final safety check happens once a creator decides to post [the creation from AI Alive] to their Story.''} (TikTok's Newsroom article).

\subsubsection{Takeaways}
Platforms govern their own integrated AI tools by applying rules and safeguards to both output production and distribution. This approach indicates that platforms consider themselves as co-producers of AI-generated content, necessitating both the safety of tools and control over downstream use. Compared to moderating user-generated content in general, this approach reflects heightened platform responsibility and accountability for AI tools, particularly when the tool output is seamlessly embedded into existing content workflows.

\subsection{Educating and Empowering Users When Interacting with AI-Generated Content}
\label{subsec: 5.6}
In our dataset, 17 platforms provide users with tools and resources to enhance user autonomy to distinguish and control AI-generated content in their feeds, and enable creators' critical literacy to responsibly make and share AI-generated content. We identified three primary forms of helping users to interact with AI-generated content: labeling systems to support AI-generated content disclosure, educational materials for both creators and users to improve their critical engagement with AI-generated content, and personalized recommendations and moderation to control the presence of AI-generated content in user feeds, which we present below.

\subsubsection{AI-Generated Content Labeling System and Label Design} 
In our dataset, the disclosure requirements for AI-generated content can be achieved through various methods, such as hashtags or conspicuous claims in a user's post on a platform. Many platforms have also developed a dedicated labeling system to empower both creators and viewers around AI-generated content more effectively. This approach supports compliance with the disclosure rules through an easier and more unified system, but also communicates more clearly to viewers when they encounter AI-generated content. We found eight out of 14 platforms (DeviantArt, Facebook, Instagram, Pixabay, Threads, TikTok, Vimeo, YouTube) with rules on disclosing AI-generated content also provide labeling systems for creators and design disclosure labels for viewers. For example, \textit{``Creators can disclose content as AI-generated directly on the post by adding text, a hashtag sticker, or context in the post's description. TikTok also offers labels to let viewers know when content was made using AI: Creator label: Creators apply this label to indicate that their content was completely generated or significantly edited by AI.''} (TikTok's Support Center article). 

Additionally, ten platforms (DeviantArt, Facebook, Instagram, LinkedIn, Pinterest, Threads, TikTok, Twitter/X, Vimeo, YouTube) describe their label designs for creator-labeling or auto-labeling that content is AI-generated. The information conveyed through these labels is diverse, ranging from short markers such as \textit{``AI Info''} (Meta platforms) or \textit{``Created using AI tools''} (DeviantArt), to more detailed descriptions that extend beyond simply indicating whether content is AI-made. A label pattern with concrete information is exemplified in the following statement: \textit{``LinkedIn members will start seeing a new icon that helps members trace the origin of AI-created media, including the source and history of the content, and whether it was created or edited by AI.''} (LinkedIn's Newsroom article). The placement of labels, according to the platforms that disclose this information, also varies---from being displayed directly on the content itself to appearing in the detailed information or metadata of the media, which requires an additional click to access. Some platforms adopt different label formats for AI-generated content with a more sensitive nature for users: \textit{``For most videos, a label will appear in the expanded description, but for videos that touch on more sensitive topics — like health, news, elections, or finance — we'll also show a more prominent label on the video itself.''} (YouTube's Official Blog).

\subsubsection{Educational Materials on Critical Engagement with AI-Generated Content}
Besides the policy, guidelines, and other informal documents that indicate user rules and governance actions on AI-generated content, a few platforms provide additional materials to educate users on a more critical engagement with AI-generated content. We found two platforms (Pixabay, Medium) that publish creator guidelines to encourage the mindful and responsible use of generative AI for content creation. Pixabay, for example, introduces a separate AI quality guideline, alongside their general content quality guidelines, for creators with the aim of \textit{``ensure[ing] that the quality of AI-generated work meets high standards and is user-friendly.''} Notably, StackOverflow, where posting any AI-generated content is strictly forbidden, designed a banner to remind users not to publish AI-generated content: \textit{``We will be graduating this experiment and adding functionality to enable the "AI-generated content" policy banner on Stack Overflow. While the variant group had no significant impact on answer rates, there was a reduction in posts flagged for AI-generated content.''} (StackOverflow's Announcement).

Some platforms also produce materials targeting the critical consumption of AI-generated content. We identified six platforms (Facebook, Instagram, Reddit, Threads, TikTok, Twitch) that have separate documents for users on how to distinguish AI-generated content and be aware of harmful AI-generated content, such as AI-generated scams. Pornhub, in addition, warns users when they proactively seek non-consensual deepfakes that can cause harm to others: \textit{``If users search for terms relating to non-consensual [deepfake] material, they are reminded that such material may be illegal.''} (Pornhub's Help Center article).

\subsubsection{Personalized Recommendation and Moderation Tailoring for AI-Generated Content}
We identified four platforms (DeviantArt, Medium, Pinterest, Pixiv), all of which are non-mainstream social media platforms, that provide users with tools to adjust the level of AI-generated content in their feeds. Typically, social media platforms enable users to adjust their recommendation feeds~\cite{li2025beyond} and to set up personalized moderation filters~\cite{jhaver2023personalizing} that control the content they see. These platforms extend and tailor such mechanisms to AI-generated content, thereby giving viewers more autonomy in interacting with AI-generated content. For example, Pinterest mentions their plans to integrate user controls on recommendation feeds specifically on AI-generated content: \textit{``We will soon launch an experiment allowing users to select a `see fewer' option on Gen AI Pins for certain categories prone to AI modification or generation, such as beauty and art and will continue expanding into more areas. [...] This will send a signal to our systems to recommend less of this type of content.''} (Pinterest's Newsroom article). Pixiv, on the other hand, employs a content filter for users to remove any AI-generated content from their feed: \textit{``You can choose whether or not to show AI-generated work in your search results. There are three options for filtering AI-generated work: `Display', `Hide'.''} (Pixiv's Help Center article). 

\subsubsection{Takeaways}
Platforms provide labeling systems, educational resources, and feed controls around AI-generated content to support user awareness and autonomy. Rather than solely relying on enforcement over users and content, these measures position user control and literacy as complementary governance tools for AI-generated content. Notably, the uneven availability of different tools and resources also highlights disparities in how much responsibility platforms delegate to users around AI-generated content.
\section{Discussion and Future Work}

From our case study of AI-generated content governance on social media platforms, we find that many platforms apply existing content moderation strategies to AI-generated content, including moderating output from integrated AI tools (Section \ref{subsec: 5.1}, Section \ref{subsec: 5.5}). However, some platforms have introduced new strategies, such as disclosure and labeling requirements and user-focused tools and educational materials (Sections \ref{subsec: 5.2}---\ref{subsec: 5.6}). Overall, platforms tend to structure their AI governance policies similarly to their general content moderation policies, covering rule definitions, detection, and enforcement.

Our findings also reveal significant gaps in current governance practices of online platforms like social media. Below, we outline recommendations and future directions for stakeholders seeking to improve the governance of AI-generated content.

\subsection{Recommendations for Platforms}
Our findings reveal unresolved questions around disclosure standards, detection reliability, and quality assessment. We also found a lack of structured policy documents and user resources. Below, we detail these gaps and offer design suggestions for platforms. We note that even on platforms that partly outsource content authenticity assessment to users, such as Twitter/X which heavily relies on Community Notes for user-driven fact-checking~\cite{prollochs2022community}, platforms are still responsible for managing and governing the infrastructural conditions under which AI-generated content is surfaced. Our recommendations, therefore, target these enduring responsibilities.

\subsubsection{Reconsider Disclosure Patterns for AI-Generated Content}
Many social media platforms have established disclosure rules for AI-generated content, but their approaches vary widely: some impose no requirements, others mandate disclosure of all AI-generated content, and still others require disclosure only for realistic media or to denote authorship (Section \ref{subsec: 5.2}). Label designs and placements also differ across platforms (Section \ref{subsec: 5.6}). Researchers have investigated user perspectives on which AI content should be labeled, how to design labels, and how labels affect users~\cite{burrus2024unmasking, gamage2025labeling, jung2025ai, Wittenberg2024Labeling, epstein2023label}. Yet these inconsistencies in practice reflect deeper unresolved questions about disclosure standards. Moreover, as AI becomes increasingly integrated into content creation, creators' use of AI spans a spectrum rather than a binary distinction~\cite{kim2024unlocking, are2025content}. Social media platforms we studied, however, only impose disclosure on whether the content involves AI-generated components or not. In fact, only a few online platforms have begun treating disclosure as non-binary. For example, Spotify has introduced a multi-layer disclosure policy covering vocals, instrumentals, and post-production, respectively, to reflect AI use across the entire music creation workflow~\cite{spotify_2025}.

To improve disclosure practices for AI-generated content, we suggest that online platforms carefully consider what types of AI-generated content should be disclosed in what way. Specifically, platforms should adopt strategies appropriate for the nature of their platforms. For example, platforms hosting videos may require labels for the visual frame and the audio track, respectively, when either is AI-generated. On music streaming platforms, disclosure could vary to indicate which parts of a song and related materials, such as lyrics, samples, and album art, are AI-generated. For platforms centered on creative works, we further recommend that AI disclosure labels be designed more prominently than on other platforms, to provide better transparency around content ownership and authorship.

\subsubsection{Navigate Unreliable Detection of AI-Generated Content}
Detecting AI-generated content remains a significant barrier to effective governance. Many social media platforms in our study rely primarily on automated detection (Section \ref{subsec: 5.2}), but algorithmic classifiers are often unreliable, exhibiting low accuracy and bias~\cite{boutadjine2025human}. Even industry-standard provenance techniques, watermarking, and content credentials can be circumvented by bad actors~\cite{jiang2023evading} and do not help detect content from unregulated open-source models. A recent study found that platforms often fail to label AI-generated content even when it carries industry-standard provenance, contradicting their stated policies~\cite{mantzarlis_dutta_2025}. While platforms could increase human review, this raises scalability concerns, and humans are not necessarily reliable arbiters of AI-generated content either~\cite{zhou2023synthetic, mink2023deepphish, boutadjine2025human, ha2024organic}.

Given these detection limitations, we suggest that platforms establish clear accountability mechanisms for any governance that relies on detecting AI-generated content. Users should be informed why and how their content was flagged as AI-generated and given a channel to provide feedback. In particular, since our findings show that appeal pipelines for automatic AI-generated content labeling are currently rare, we recommend that platforms provide mechanisms for such user appeal. Users should be able to contest when their content is incorrectly labeled as AI-generated, similar to existing appeals for content removals.

\subsubsection{Extend Governance Actions to Quality, Ownership, and Authorship}
Our findings reveal a general lack of governance around quality, ownership, and authorship of AI-generated content on social media. Only a few platforms focused on knowledge sharing and content creation have AI-specific restrictions on ownership or the value placed on AI-generated content (Section \ref{subsec: 5.3}). Similarly, few platforms restrict creators from monetizing AI-generated content (Section \ref{subsec: 5.4}). Growing public debates about ``AI slop'' (low-quality AI-generated media)~\cite{mahdawi2025slop}, AI-generated spam~\cite{diresta2024spammers}, and ownership and credit for AI-generated content~\cite{lima2025public, 10.1145/3706598.3713522, moller2025impact, wei2024pixiv} reflect an urgent need for governance action. Content quality, ownership, and authorship are also major concerns for creators, who increasingly worry about their rights and creative identity~\cite{kyi2025governance, gero2025creative, shan2023glaze, shan2024nightshade, are2025content}. 

Given the increasing use of AI in content creation and mixed reactions from users and creators, we recommend that platforms explicitly address quality, ownership, and authorship of AI-generated content. Doing so is essential for safeguarding community values and protecting creator rights. Platforms that allow monetization should evaluate creator eligibility to prevent profit from low-effort AI-generated content. Specifically, platforms can take action by developing clear and concrete guidelines for creators on using AI to create high-quality content that aligns with community values. Since the quality, ownership, and authorship of AI-generated content are difficult to define objectively, platforms could also incorporate user-driven mechanisms, such as community rating systems or user-driven flagging, similar to community notes on Twitter/X for crowdsourced fact-checking~\cite{prollochs2022community}, to help surface low-value AI-generated content.



\subsubsection{Inform Users About Governance Actions Through Consistent, Explicit Communication}

About two-thirds of platforms in our study explicitly reference their governance of AI-generated content, but most simply describe how existing moderation policies apply rather than establishing new AI-specific policies (Section \ref{subsec: 5.1}). The remaining platforms have no explicit references and rely implicitly on existing policies. Although this reflects an evolving ecosystem, the increasing risks of AI-generated content (e.g., deepfake political figures~\cite{lundberg2025potential, gamage2022deepfakes}) necessitate more direct communication with users about governance, even when existing policies apply.

Even platforms that explicitly addressed AI governance had this information scattered across different page types (Section \ref{sec:data}). This echoes prior observations that content moderation policies are often dispersed, creating challenges for users and regulators alike~\cite{schaffner2024community, gao2025cannot}. Unlike platforms with community guidelines pages consolidating moderation information~\cite{schaffner2024community, jiang2020characterizing}, all platforms in our study except four lacked a consolidated page summarizing AI-generated content policies (Section \ref{sec:data}). 

To address these gaps, we recommend that platforms make their AI governance more explicit and accessible. Platforms should create dedicated policy pages outlining governance rationale, rules, detection methods, and enforcement procedures. Similar to how content moderation policies use concrete examples~\cite{schaffner2024community}, platforms could illustrate what counts as AI-generated content and what types are allowed or restricted. Platforms should also create a meta-policy page summarizing key governance actions with links to relevant documents, similar to Medium’s dedicated AI Content Policy page.\footnote{\url{https://help.medium.com/hc/en-us/articles/22576852947223-Artificial-Intelligence-AI-content-policy}} By consolidating AI governance information, platforms can reduce ambiguity and help users understand how AI-generated content fits into broader trust and safety expectations.

\subsubsection{Develop More Educational Materials and Controls for Users}

Many social media platforms rely on labels to inform users that content is AI-generated, but few offer educational materials to help users recognize such content and its risks (Section \ref{subsec: 5.6}). Automated detection and labeling systems remain inaccurate~\cite{boutadjine2025human, mantzarlis_dutta_2025}, and humans increasingly struggle to distinguish AI-generated content~\cite{zhou2023synthetic, mink2023deepphish, boutadjine2025human, ha2024organic}. As a result, labels alone may not prevent users from being misled---and may even deepen confusion. Educational resources are therefore essential for helping users proactively recognize AI-generated content and its risks, particularly when moderation or labeling fails. This is especially important for populations with limited digital literacy, such as older adults and children~\cite{lao2025everyday}. Platforms should incorporate AI-generated content into their media literacy materials, with concrete examples of prevalence, risks, and detection principles.

Similarly, few platforms provide features for users to control AI-generated content in their feeds (Section \ref{subsec: 5.6}). While many users enjoy viewing and engaging with AI-generated content, others are concerned about the diminishing value of human-made content and degraded experience quality due to idea convergence in generative AI~\cite{moller2025impact}. Given these varied perspectives, platforms should offer users autonomy over their exposure to AI-generated content to prevent distraction and mental burden from unwillingness to view it. Many platforms already allow feed customization through filters and recommendation adjustments~\cite{jhaver2023personalizing, li2025beyond}, but these typically focus on content topics rather than production methods. We recommend that platforms implement similar feed controls tailored specifically to AI-generated content---allowing users to choose whether to see such content and, if so, to what degree they wish to be exposed to it.

\subsection{Recommendations for Regulators}
As with content moderation~\cite{schaffner2024community}, the governance of AI-generated content is informed by regulations but dominated by platforms themselves. Current regulations primarily focus on mitigating direct harms, such as defining liability of AI service providers and penalizing distributors of non-consensual deepfakes. Based on the gaps we identified, we outline directions for future regulatory development.

\subsubsection{Update Existing Laws for AI-Generated Content}
Platforms rely heavily on existing laws to handle content violations such as copyright infringement and online scams. However, these laws often fail to address AI-generated content due to outdated definitions, leaving platforms without clear guidance. For example, few social media platforms we studied have policies on ownership and authorship of AI-generated content (Section \ref{subsec: 5.3}), likely because copyright law lacks clear definitions for how training data and outputs should be treated~\cite{samuelson2023generative}. This ambiguity also affects monetization, as platforms struggle to determine who can profit from AI-generated content without clear copyright boundaries. Consequently, harms like mass production of low-quality AI spam for monetization cannot be addressed through existing copyright or deepfake laws alone.


We recommend that regulators update legal frameworks to explicitly address ownership, authorship, liability, and commercialization of AI-generated content. For example, Huang et al. proposed defining copyright for AI-generated content by modeling the relationship between output uncertainty and the degree of human creative control over the generation process~\cite{10.1145/3715336.3735683}. If adopted, such a framework could quantify human creators' contributions and guide platform monetization policies.


\subsubsection{Develop Standardized Guidelines for AI-Generated Content Governance}
In our study, we found governance of AI-generated content varies substantially across platforms, both in policy structure and in specific mechanisms like disclosure systems and labels for AI-generated content. Prior work has found that AI developers often lack full awareness of negative societal impacts, such as non-consensual deepfakes and privacy violations. This limited awareness, combined with business incentives, creates blind spots in platform-driven governance and underscores the need for regulatory oversight~\cite{pawelec2025decent}.

At this early stage, regulators should develop official guidelines and standardized frameworks for platforms. These could function similarly to government guidance for enterprise digital security controls~\cite{ruth2025first} or the minimum obligations, transparency requirements, and user protections mandated by the EU Digital Services Act~\cite{eu_dsa}. Such guidelines should offer coherent requirements and best practices, including mandating explicit AI policies, clarifying provenance expectations, and specifying user protection requirements.

\subsection{Future Work for HCI Researchers}
Our study provides a baseline of current AI-generated content governance on major social media platforms. We suggest several directions for future HCI research.

\subsubsection{Longitudinal Research on Governance Changes}
Our findings on the governance of AI-generated content on social media are based on one-time
snapshot at the time of collection. However, as generative AI evolves rapidly, governance practices will likely undergo continuous change. Future work could examine how platform policies and enforcement shift over time. Longitudinal studies, including repeated audits and time-series analyses, would help capture governance adaptation patterns and the impact of external events such as product releases, regulatory updates, and public incidents.

\subsubsection{Investigating and Improving Governance Mechanisms}
Our findings identify governance mechanisms needing further investigation, including quality standards, disclosure labels, educational materials, and user controls for filtering AI-generated content. Future work could evaluate these mechanisms on real platforms---for example, measuring disclosure practices, investigating user perceptions of quality and monetization, and testing educational interventions. Researchers could also design novel governance tools, such as disclosure systems or adaptive user controls. 

\subsubsection{In-depth Studies of Specific Platform Types}
While our cross-platform analysis identifies six governance approaches, further research is needed to understand how these approaches manifest, diverge, and fail within individual platforms. Such studies are necessary to tailor governance to platform-specific needs---for instance, distinguishing governance requirements between social networks and music platforms. Researchers should engage diverse stakeholders, including viewers, traditional creators, AI content creators, and advertisers, to understand how each group perceives and is affected by AI-generated content governance within specific platform ecosystems.

\section{Conclusion}

As AI-generated content becomes increasingly prevalent on online platforms, it is
critical to understand how platforms govern this content. This paper examines
how 40 popular social media platforms have adapted their policies to address new
challenges posed by AI-generated content. Our analysis reveals that two-thirds
of these platforms explicitly articulate governance approaches for AI-generated
content, spanning six distinct categories of action. The most common strategies
involve extending existing content moderation frameworks to encompass
inappropriate AI-generated content, as well as articulating disclosure and
labeling requirements. Platforms primarily focused on creativity and
knowledge-sharing typically have more comprehensive policies, often addressing
more nuanced issues of quality, authorship, and ownership. Platforms that have
integrated AI tools tend to prioritize safety measures for content generation
and distribution. Finally, some platforms empower users through labeling
mechanisms, educational resources about AI-generated content risks, and
personalized controls for managing such content in their feeds.
Our findings reveal significant variation in how platforms approach AI-generated content governance, with many relying on existing policies rather than developing comprehensive AI-specific frameworks. We recommend that platforms adopt more explicit, unified, and forward-looking governance strategies that directly address the unique challenges posed by AI-generated content. 

\begin{acks}
This work was funded by Google.
\end{acks}

\bibliographystyle{ACM-Reference-Format}
\bibliography{ref}

@inproceedings{singhal2023sok,
  title={SoK: Content moderation in social media, from guidelines to enforcement, and research to practice},
  author={Singhal, Mohit and Ling, Chen and Paudel, Pujan and Thota, Poojitha and Kumarswamy, Nihal and Stringhini, Gianluca and Nilizadeh, Shirin},
  booktitle={2023 IEEE 8th European Symposium on Security and Privacy (EuroS\&P)},
  pages={868--895},
  year={2023},
  publisher={IEEE},
  doi={10.1109/EuroSP57164.2023.00056}
}

@inproceedings{schaffner2024community,
  title={``Community guidelines make this the best party on the internet'': an in-depth study of online platforms' content moderation policies},
  author={Schaffner, Brennan and Bhagoji, Arjun Nitin and Cheng, Siyuan and Mei, Jacqueline and Shen, Jay L and Wang, Grace and Chetty, Marshini and Feamster, Nick and Lakier, Genevieve and Tan, Chenhao},
  booktitle={Proceedings of the 2024 CHI Conference on Human Factors in Computing Systems},
  pages={1--16},
  year={2024},
  publisher={ACM},
  doi={10.1145/3613904.3642333}
}

@article{fisher2024moderating,
  title={Moderating synthetic content: The challenge of generative AI},
  author={Fisher, Sarah A and Howard, Jeffrey W and Kira, Beatriz},
  journal={Philosophy \& Technology},
  volume={37},
  number={4},
  pages={133},
  year={2024},
  publisher={Springer},
  doi={https://doi.org/10.1007/s13347-024-00818-9}
}

@inproceedings{jiang2020characterizing,
  title={Characterizing community guidelines on social media platforms},
  author={Jiang, Jialun `Aaron' and Middler, Skyler and Brubaker, Jed R and Fiesler, Casey},
  booktitle={Companion Publication of the 2020 Conference on Computer Supported Cooperative Work and Social Computing},
  pages={287--291},
  year={2020},
  publisher={ACM},
  doi={https://doi.org/10.1145/3406865.3418312}
}

@article{arora2023detecting,
  title={Detecting harmful content on online platforms: what platforms need vs. where research efforts go},
  author={Arora, Arnav and Nakov, Preslav and Hardalov, Momchil and Sarwar, Sheikh Muhammad and Nayak, Vibha and Dinkov, Yoan and Zlatkova, Dimitrina and Dent, Kyle and Bhatawdekar, Ameya and Bouchard, Guillaume and others},
  journal={ACM Computing Surveys},
  volume={56},
  number={3},
  pages={1--17},
  year={2023},
  publisher={ACM New York, NY},
  doi={10.1145/3603399}
}

@article{moller2025impact,
  title={The Impact of Generative AI on Social Media: An Experimental Study},
  author={M{\o}ller, Anders Giovanni and Romero, Daniel M and Jurgens, David and Aiello, Luca Maria},
  journal={arXiv preprint arXiv:2506.14295},
  year={2025},
  doi={https://doi.org/10.48550/arXiv.2506.14295}
}

@article{kietzmann2011social,
  title={Social media? Get serious! Understanding the functional building blocks of social media},
  author={Kietzmann, Jan H and Hermkens, Kristopher and McCarthy, Ian P and Silvestre, Bruno S},
  journal={Business horizons},
  volume={54},
  number={3},
  pages={241--251},
  year={2011},
  publisher={Elsevier},
  doi={https://doi.org/10.1016/j.bushor.2011.01.005}
}

@inproceedings{lloyd2025ai,
  title={AI Rules? Characterizing Reddit Community Policies Towards AI-Generated Content},
  author={Lloyd, Travis and Gosciak, Jennah and Nguyen, Tung and Naaman, Mor},
  booktitle={Proceedings of the 2025 CHI Conference on Human Factors in Computing Systems},
  pages={1--19},
  year={2025},
  doi={https://doi.org/10.1145/3706598.3713292}
}

@article{lloyd2023there,
  title={``There Has To Be a Lot That We're Missing'': Moderating AI-Generated Content on Reddit},
  author={Lloyd, Travis and Reagle, Joseph and Naaman, Mor},
  journal={Proceedings of the ACM on Human-Computer Interaction (CSCW)},
  year={2025},
  doi={https://dl.acm.org/doi/10.1145/3757445}
}

@article{schaffner2022understanding,
author = {Schaffner, Brennan and Lingareddy, Neha A. and Chetty, Marshini},
title = {Understanding Account Deletion and Relevant Dark Patterns on Social Media},
year = {2022},
issue_date = {November 2022},
publisher = {Association for Computing Machinery},
address = {New York, NY, USA},
volume = {6},
number = {CSCW2},
url = {https://doi.org/10.1145/3555142},
doi = {10.1145/3555142},
journal = {Proc. ACM Hum.-Comput. Interact.},
month = nov,
articleno = {417},
numpages = {43}
}

@inproceedings{gao2025cannot,
  title={``I Cannot Write This Because It Violates Our Content Policy'': Understanding Content Moderation Policies and User Experiences in Generative AI Products},
  author={Gao, Lan and Chen, Oscar and Lee, Rachel and Feamster, Nick and Tan, Chenhao and Chetty, Marshini},
  booktitle={34th USENIX Security Symposium (USENIX Security 25)},
  pages={3727--3746},
  year={2025},
  url={https://www.usenix.org/conference/usenixsecurity25/presentation/gao-lan}
}

@inproceedings{pater2016characterizations,
author = {Pater, Jessica A. and Kim, Moon K. and Mynatt, Elizabeth D. and Fiesler, Casey},
title = {Characterizations of Online Harassment: Comparing Policies Across Social Media Platforms},
year = {2016},
isbn = {9781450342766},
publisher = {Association for Computing Machinery},
address = {New York, NY, USA},
url = {https://doi.org/10.1145/2957276.2957297},
doi = {10.1145/2957276.2957297},
booktitle = {Proceedings of the 2016 ACM International Conference on Supporting Group Work},
pages = {369–374},
numpages = {6},
location = {Sanibel Island, Florida, USA},
series = {GROUP '16}
}

@article{almansoori2025can,
  title={Can Social Media Privacy and Safety Features Protect Targets of Interpersonal Attacks? A Systematic Analysis},
  author={Almansoori, Majed and Chatterjee, Rahul},
  journal={Proceedings on Privacy Enhancing Technologies},
  year={2025},
  doi={https://doi.org/10.56553/popets-2025-0064}
}

@article{sun2025aigt,
  title={Are we in the AI-generated text world already? Quantifying and monitoring AIGT on social media},
  author={Sun, Zhen and Zhang, Zongmin and Shen, Xinyue and Zhang, Ziyi and Liu, Yule and Backes, Michael and Zhang, Yang and He, Xinlei},
  booktitle={Proceedings of the 63rd Annual Meeting of the Association for Computational Linguistics (Volume 1: Long Papers)},
  pages={22975--23005},
  year={2025},
  doi ={10.18653/v1/2025.acl-long.1120}
}

@article{matatov2024subreddits,
  title={Examining the prevalence and dynamics of AI-generated media in art subreddits},
  author={Matatov, Hana and Qu{\'e}r{\'e}, Marianne Aubin Le and Amir, Ofra and Naaman, Mor},
  journal={arXiv preprint arXiv:2410.07302},
  year={2024},
  doi={https://doi.org/10.48550/arXiv.2410.07302}
}

@inproceedings{wei2024pixiv,
author = {Wei, Yiluo and Tyson, Gareth},
title = {Understanding the Impact of AI-Generated Content on Social Media: The Pixiv Case},
year = {2024},
isbn = {9798400706868},
publisher = {Association for Computing Machinery},
address = {New York, NY, USA},
url = {https://doi.org/10.1145/3664647.3680631},
doi = {10.1145/3664647.3680631},
booktitle = {Proceedings of the 32nd ACM International Conference on Multimedia},
pages = {6813–6822},
numpages = {10},
location = {Melbourne VIC, Australia},
series = {MM '24}
}

@inproceedings{lyu2024youtube,
author = {Lyu, Yao and Zhang, He and Niu, Shuo and Cai, Jie},
title = {A Preliminary Exploration of YouTubers' Use of Generative-AI in Content Creation},
year = {2024},
isbn = {9798400703317},
publisher = {Association for Computing Machinery},
address = {New York, NY, USA},
url = {https://doi.org/10.1145/3613905.3651057},
doi = {10.1145/3613905.3651057},
booktitle = {Extended Abstracts of the CHI Conference on Human Factors in Computing Systems},
articleno = {20},
numpages = {7},
location = {Honolulu, HI, USA},
series = {CHI EA '24}
}

@article{dmonte2024electionclaims,
  title={Classifying human-generated and ai-generated election claims in social media},
  author={Dmonte, Alphaeus and Zampieri, Marcos and Lybarger, Kevin and Albanese, Massimiliano and Coulter, Genya},
  journal={arXiv preprint arXiv:2404.16116},
  year={2024},
  doi={https://doi.org/10.48550/arXiv.2404.16116}
}

@inproceedings{mink2023deepphish,
author = {Jaron Mink and Licheng Luo and Nat{\~a} M. Barbosa and Olivia Figueira and Yang Wang and Gang Wang},
title = {DeepPhish: Understanding User Trust Towards Artificially Generated Profiles in Online Social Networks},
booktitle = {31st USENIX Security Symposium (USENIX Security 22)},
year = {2022},
pages = {1669--1686},
url = {https://www.usenix.org/conference/usenixsecurity22/presentation/mink}
}

@article{qiwei2024deepfakes,
  title={Reporting non-consensual intimate media: An audit study of deepfakes},
  author={Qiwei, Li and Zhang, Shihui and Kasper, Andrew Timothy and Ashkinaze, Joshua and Eaton, Asia A and Schoenebeck, Sarita and Gilbert, Eric},
  journal={arXiv preprint arXiv:2409.12138},
  year={2024},
  doi={https://doi.org/10.48550/arXiv.2409.12138}
}

@article{ma2022fairness,
  author    = {Ma, Renkai and Kou, Yubo},
  title     = {``I'm Not Sure What Difference Is Between Their Content and Mine, Other Than the Person Itself'': A Study of Fairness Perception of Content Moderation on YouTube},
  journal   = {Proceedings of the ACM on Human-Computer Interaction},
  volume    = {6},
  number    = {CSCW},
  pages     = {425:1--425:28},
  year      = {2022},
  publisher = {ACM},
  address   = {New York, NY, USA},
  doi       = {10.1145/3555150},
  url       = {https://doi.org/10.1145/3555150}
}

@article{myerswest2018censored,
  author    = {Myers West, Sarah},
  title     = {Censored, Suspended, Shadowbanned: User Interpretations of Content Moderation on Social Media Platforms},
  journal   = {New Media \& Society},
  volume    = {20},
  number    = {11},
  pages     = {4366--4383},
  year      = {2018},
  publisher = {SAGE Publications},
  doi       = {10.1177/1461444818773059},
  url       = {https://doi.org/10.1177/1461444818773059}
}

@article{gorwa2019platform,
  author    = {Gorwa, Robert},
  title     = {What is Platform Governance?},
  journal   = {Information, Communication \& Society},
  volume    = {22},
  number    = {6},
  pages     = {854--871},
  year      = {2019},
  publisher = {Taylor \& Francis},
  doi       = {10.1080/1369118X.2019.1573914},
  url       = {https://doi.org/10.1080/1369118X.2019.1573914}
}

@inproceedings{obrien2025dating,
author = {O'Brien, Catherine R K and Roslan, Nuur Alifah and Murdoch, Steven J. and Abu-Salma, Ruba and Zytko, Douglas and Warner, Mark},
title = {Online Dating Platform Safeguards and Self-Protection: How Dating Platforms Characterise, Respond to, and Safeguard Against Harms},
year = {2025},
isbn = {9798400713958},
publisher = {Association for Computing Machinery},
address = {New York, NY, USA},
url = {https://doi.org/10.1145/3706599.3719825},
doi = {10.1145/3706599.3719825},
booktitle = {Proceedings of the Extended Abstracts of the CHI Conference on Human Factors in Computing Systems},
articleno = {423},
numpages = {8},
location = {
},
series = {CHI EA '25}
}

@article{feng2023examining,
  title={Examining the impact of provenance-enabled media on trust and accuracy perceptions},
  author={Feng, KJ Kevin and Ritchie, Nick and Blumenthal, Pia and Parsons, Andy and Zhang, Amy X},
  journal={Proceedings of the ACM on Human-Computer Interaction},
  volume={7},
  number={CSCW2},
  pages={1--42},
  year={2023},
  publisher={ACM New York, NY, USA},
  doi={https://doi.org/10.1145/3610061}
}

@article{Wittenberg2024Labeling,
	author = {Wittenberg, Chloe and Epstein, Ziv and Berinsky, Adam J. and Rand, David G.},
	journal = {An MIT Exploration of Generative AI},
	year = {2024},
	url = {https://mit-genai.pubpub.org/pub/hu71se89},
	publisher = {MIT},
	title = {Labeling {AI}-{Generated} {Content}: Promises, {Perils}, and {Future} {Directions}},
}

@article{burrus2024unmasking,
  title={Unmasking AI: Informing authenticity decisions by labeling AI-generated content},
  author={Burrus, Olivia and Curtis, Amanda and Herman, Laura},
  journal={Interactions},
  volume={31},
  number={4},
  pages={38--42},
  year={2024},
  publisher={ACM New York, NY, USA},
  doi={https://doi.org/10.1145/3665321}
}

@inproceedings{gamage2025labeling,
  title={Labeling Synthetic Content: User Perceptions of Label Designs for AI-Generated Content on Social Media},
  author={Gamage, Dilrukshi and Sewwandi, Dilki and Zhang, Min and Bandara, Arosha K},
  booktitle={Proceedings of the 2025 CHI Conference on Human Factors in Computing Systems},
  pages={1--29},
  year={2025},
  doi={https://doi.org/10.1145/3706598.3713171}
}

@inproceedings{jung2025ai,
  title={AI-Generated or AI-Modified? User Reactions to Labeling AI Use in Social Media Posts},
  author={Jung, Yongnam and Hua, Peixin and Bao, Jiaqi and Sundar, S Shyam},
  booktitle={Proceedings of the Extended Abstracts of the CHI Conference on Human Factors in Computing Systems},
  pages={1--7},
  year={2025},
  doi={https://doi.org/10.1145/3706599.3720264}
}

@article{epstein2023label,
  title={What label should be applied to content produced by generative AI},
  author={Epstein, Ziv and Fang, Mengying Cathy and Arechar, Antonio Alonso and Rand, David G},
  journal={PsyArXiv},
  volume={10},
  pages={10--31234},
  year={2023},
  doi={https://doi.org/10.31234/osf.io/v4mfz}
}

@inproceedings{hua2024generative,
  title={Generative AI in user-generated content},
  author={Hua, Yiqing and Niu, Shuo and Cai, Jie and Chilton, Lydia B and Heuer, Hendrik and Wohn, Donghee Yvette},
  booktitle={Extended Abstracts of the CHI Conference on Human Factors in Computing Systems},
  pages={1--7},
  year={2024},
  doi={https://doi.org/10.1145/3613905.3636315}
}

@article{burtch2023consequences,
  title={The consequences of generative AI for UGC and online community engagement},
  author={Burtch, Gordon and Lee, Dokyun and Chen, Zhichen},
  journal={Available at SSRN 4521754},
  year={2023},
  url={https://papers.ssrn.com/sol3/papers.cfm?abstract_id=4521754}
}

@inproceedings{kim2024unlocking,
  title={Unlocking creator-AI synergy: Challenges, requirements, and design opportunities in AI-powered short-form video production},
  author={Kim, Jini and Kim, Hajun},
  booktitle={Proceedings of the 2024 CHI Conference on Human Factors in Computing Systems},
  pages={1--23},
  year={2024},
  doi={https://doi.org/10.1145/3613904.3642476}
}

@article{chen2025synthetic,
  title={Synthetic politics: Prevalence, spreaders, and emotional reception of AI-generated political images on X},
  author={Chen, Zhiyi and Ye, Jinyi and Tsai, Beverlyn and Ferrara, Emilio and Luceri, Luca},
  journal={arXiv preprint arXiv:2502.11248},
  year={2025},
  doi={https://doi.org/10.48550/arXiv.2502.11248}
}

@article{shahid2024examining,
  title={Examining Human-AI Collaboration for Co-Writing Constructive Comments Online},
  author={Shahid, Farhana and Dittgen, Maximilian and Naaman, Mor and Vashistha, Aditya},
  journal={Proc. ACM Hum.-Comput. Interact. 9, 7, Article 410 (November 2025)},
  year={2025},
  doi={https://doi.org/10.1145/3757591}
}

@article{ma2025social,
  title={Social, legal, and ethical implications of AI-Generated deepfake pornography on digital platforms: A systematic literature review},
  author={Furizal and Ma'arif, Alfian and Maghfiroh, Hari and Suwarno, Iswanto and Prayogi, Denis and Kariyamin and Lonang, Syahrani and Sharkawy, Abdel-Nasser},
  journal={Social Sciences \& Humanities Open},
  volume={12},
  pages={101882},
  year={2025},
  publisher={Elsevier},
  doi={https://doi.org/10.1016/j.ssaho.2025.101882}
}

@misc{C2PA2025,
  title        = {Coalition for Content Provenance and Authenticity (C2PA)},
  author       = {{Coalition for Content Provenance and Authenticity}},
  howpublished = {\url{https://c2pa.org/}},
  note         = {Accessed: 2025-09-01},
  year         = {2025}
}

@article{gao2025does,
  title={Does Social Bot Help Socialize? Evidence from a Microblogging Platform},
  author={Gao, Yang and Zhang, Maggie Mengqing and Lysyakov, Mikhail},
  journal={Information Systems Research},
  year={2025},
  publisher={INFORMS},
  doi={https://doi.org/10.1287/isre.2024.1089}
}

@article{diresta2024spammers,
  title={How spammers and scammers leverage AI-generated images on Facebook for audience growth},
  author={DiResta, Renee and Goldstein, Josh A},
  journal={arXiv preprint arXiv:2403.12838},
  year={2024},
  doi={https://doi.org/10.48550/arXiv.2403.12838}
}

@inproceedings{mink2024s,
  title={It's Trying Too Hard To Look Real: Deepfake Moderation Mistakes and Identity-Based Bias},
  author={Mink, Jaron and Wei, Miranda and Munyendo, Collins W and Hugenberg, Kurt and Kohno, Tadayoshi and Redmiles, Elissa M and Wang, Gang},
  booktitle={Proceedings of the 2024 CHI Conference on Human Factors in Computing Systems},
  pages={1--20},
  year={2024},
  doi={https://doi.org/10.1145/3613904.3641999}
}

@article{kreps2022all,
  title={All the news that’s fit to fabricate: AI-generated text as a tool of media misinformation},
  author={Kreps, Sarah and McCain, R Miles and Brundage, Miles},
  journal={Journal of experimental political science},
  volume={9},
  number={1},
  pages={104--117},
  year={2022},
  publisher={Cambridge University Press},
  doi={10.1017/XPS.2020.37}
}

@inproceedings{rae2024effects,
  title={The effects of perceived AI use on content perceptions},
  author={Rae, Irene},
  booktitle={Proceedings of the 2024 CHI Conference on Human Factors in Computing Systems},
  pages={1--14},
  year={2024},
  doi={https://doi.org/10.1145/3613904.3642076}
}

@article{kiesler2012regulating,
  title={Regulating behavior in online communities},
  author={Kiesler, Sara and Kraut, Robert and Resnick, Paul and Kittur, Aniket},
  journal={Building successful online communities: Evidence-based social design},
  volume={1},
  pages={4--2},
  year={2012},
  publisher={MIT Press Cambridge, MA},
  doi={https://doi.org/10.7551/mitpress/8472.003.0005}
}

@article{buckley2022censorship,
  title={`Censorship-free'platforms: Evaluating content moderation policies and practices of alternative social media},
  author={Buckley, Nicole and Schafer, Joseph S},
  year={2022},
  journal = {For(e)Dialogue},
  number = {Vol 4, Issue 1},
  doi={https://doi.org/10.21428/e3990ae6.483f18da}
}

@inproceedings{fiesler2018reddit,
  title={Reddit rules! characterizing an ecosystem of governance},
  author={Fiesler, Casey and Jiang, Jialun and McCann, Joshua and Frye, Kyle and Brubaker, Jed},
  booktitle={Proceedings of the International AAAI Conference on Web and Social Media},
  volume={12},
  number={1},
  pages={302--311},
  year={2018},
  publisher={Association for the Advancement of Artificial Intelligence},
  doi={https://doi.org/10.1609/icwsm.v12i1.15033}
}

@article{ma2023users,
  title={How do users experience moderation?: A systematic literature review},
  author={Ma, Renkai and You, Yue and Gui, Xinning and Kou, Yubo},
  journal={Proceedings of the ACM on Human-Computer Interaction},
  volume={7},
  number={CSCW2},
  pages={1--30},
  year={2023},
  publisher={ACM New York, NY, USA},
  doi={https://doi.org/10.1145/3610069}
}

@inproceedings{lyons2022s,
  title={What’s the appeal? Perceptions of review processes for algorithmic decisions},
  author={Lyons, Henrietta and Wijenayake, Senuri and Miller, Tim and Velloso, Eduardo},
  booktitle={Proceedings of the 2022 CHI Conference on Human Factors in Computing Systems},
  pages={1--15},
  year={2022},
  doi={https://doi.org/10.1145/3491102.3517606}
}

@inproceedings{fiesler2016reality,
  title={Reality and perception of copyright terms of service for online content creation},
  author={Fiesler, Casey and Lampe, Cliff and Bruckman, Amy S},
  booktitle={Proceedings of the 19th ACM conference on computer-supported cooperative work \& social computing},
  pages={1450--1461},
  year={2016},
  doi={https://doi.org/10.1145/2818048.2819931}
}

@article{jhaver2019did,
  title={``Did you suspect the post would be removed?'' Understanding user reactions to content removals on Reddit},
  author={Jhaver, Shagun and Appling, Darren Scott and Gilbert, Eric and Bruckman, Amy},
  journal={Proceedings of the ACM on human-computer interaction},
  volume={3},
  number={CSCW},
  pages={1--33},
  year={2019},
  publisher={ACM New York, NY, USA},
  doi={https://doi.org/10.1145/3359294}
}

@article{chandrasekharan2017you,
  title={You can't stay here: The efficacy of reddit's 2015 ban examined through hate speech},
  author={Chandrasekharan, Eshwar and Pavalanathan, Umashanthi and Srinivasan, Anirudh and Glynn, Adam and Eisenstein, Jacob and Gilbert, Eric},
  journal={Proceedings of the ACM on human-computer interaction},
  volume={1},
  number={CSCW},
  pages={1--22},
  year={2017},
  publisher={ACM New York, NY, USA},
  doi={https://doi.org/10.1145/3134666}
}

@article{jhaver2019does,
  title={Does transparency in moderation really matter? User behavior after content removal explanations on reddit},
  author={Jhaver, Shagun and Bruckman, Amy and Gilbert, Eric},
  journal={Proceedings of the ACM on Human-Computer Interaction},
  volume={3},
  number={CSCW},
  pages={1--27},
  year={2019},
  publisher={ACM New York, NY, USA},
  doi={https://doi.org/10.1145/3359252}
}

@article{vaccaro2020end,
  title={``At the End of the Day Facebook Does What It Wants'' How Users Experience Contesting Algorithmic Content Moderation},
  author={Vaccaro, Kristen and Sandvig, Christian and Karahalios, Karrie},
  journal={Proceedings of the ACM on human-computer interaction},
  volume={4},
  number={CSCW2},
  pages={1--22},
  year={2020},
  publisher={ACM New York, NY, USA},
  doi={https://doi.org/10.1145/3415238}
}

@misc{CLT2025AILabeling,
  title        = {Measures for Labeling of AI-Generated Synthetic Content},
  author       = {{China Law Translate}},
  howpublished = {\url{https://www.chinalawtranslate.com/en/ai-labeling/}},
  year         = {2025},
  note         = {Accessed: 2025-09-02},
}

@misc{Texas_SB1361_2023,
  title        = {Texas Senate Bill 1361, 88th Legislature, Regular Session},
  author       = {{Texas Legislature}},
  howpublished = {\url{https://capitol.texas.gov/tlodocs/88R/billtext/html/SB01361S.htm}},
  year         = {2023},
  note         = {Accessed: 2025-09-02},
}

@misc{Texas_SB751_2019,
  title        = {Texas Senate Bill 751, 86th Legislature, Regular Session},
  author       = {{Texas Legislature}},
  howpublished = {\url{https://capitol.texas.gov/tlodocs/86R/billtext/html/SB00751S.htm}},
  year         = {2019},
  note         = {Accessed: 2025-09-02},
}

@misc{China_GenerativeAI_Interim_2023,
  title        = {Interim Measures for the Management of Generative Artificial Intelligence Services},
  author       = {{China Law Translate}},
  howpublished = {\url{https://www.chinalawtranslate.com/en/generative-ai-interim/}},
  year         = {2023},
  note         = {Accessed: 2025-09-02},
}

@misc{EU_AIAct_Art50_2024,
  title        = {EU Artificial Intelligence Act Article 50: Transparency Obligations for Providers and Deployers of Certain AI Systems},
  author       = {{European Union}},
  howpublished = {\url{https://artificialintelligenceact.eu/article/50/}},
  year         = {2024},
  note         = {Accessed: 2025-09-02},
}

@misc{CBS_MrDeepfakes_Shutdown_2025,
  author       = {CBS News},
  title        = {AI-generated porn site Mr. Deepfakes shuts down after service provider pulls support},
  howpublished = {\url{https://www.cbsnews.com/news/ai-generated-porn-site-mr-deepfakes-shuts-down/}},
  year         = {2025},
  note         = {Accessed: 2025-09-02},
}

@misc{US_TakeItDown_2025,
  title        = {Tools to Address Known Exploitation by Immobilizing Technological Deepfakes on Websites and Networks Act (TAKE IT DOWN Act)},
  author       = {{U.S. Congress}},
  howpublished = {\url{https://www.congress.gov/bill/119th-congress/senate-bill/146}},
  year         = {2025},
  note         = {Accessed: 2025-09-02},
}

@inproceedings{matias2018civilservant,
  title={CivilServant: Community-led experiments in platform governance},
  author={Matias, J Nathan and Mou, Merry},
  booktitle={Proceedings of the 2018 CHI conference on human factors in computing systems},
  pages={1--13},
  year={2018},
  doi={https://doi.org/10.1145/3173574.3173583}
}

@inproceedings{han2025characterizing,
  title={Characterizing the MrDeepFakes Sexual Deepfake Marketplace},
  author={Han, Catherine and Li, Anne and Kumar, Deepak and Durumeric, Zakir},
  booktitle={34th USENIX Security Symposium (USENIX Security 25)},
  pages={5169--5188},
  year={2025},
  url={https://www.usenix.org/conference/usenixsecurity25/presentation/han}
}

@inproceedings{gibson2025analyzing,
  title={Analyzing the AI Nudification Application Ecosystem},
  author={Gibson, Cassidy and Olszewski, Daniel and Brigham, Natalie Grace and Crowder, Anna and Butler, Kevin RB and Traynor, Patrick and Redmiles, Elissa M and Kohno, Tadayoshi},
  booktitle={34th USENIX Security Symposium (USENIX Security 25)},
  pages={1--20},
  year={2025},
  url={https://www.usenix.org/conference/usenixsecurity25/presentation/gibson}
}

@article{goodfellow2014generative,
  title={Generative adversarial nets},
  author={Goodfellow, Ian J and Pouget-Abadie, Jean and Mirza, Mehdi and Xu, Bing and Warde-Farley, David and Ozair, Sherjil and Courville, Aaron and Bengio, Yoshua},
  journal={Advances in neural information processing systems},
  volume={27},
  year={2014},
  url={https://proceedings.neurips.cc/paper_files/paper/2014/file/f033ed80deb0234979a61f95710dbe25-Paper.pdf}
}

@inproceedings{zhou2023synthetic,
  title={Synthetic lies: Understanding ai-generated misinformation and evaluating algorithmic and human solutions},
  author={Zhou, Jiawei and Zhang, Yixuan and Luo, Qianni and Parker, Andrea G and De Choudhury, Munmun},
  booktitle={Proceedings of the 2023 CHI conference on human factors in computing systems},
  pages={1--20},
  year={2023},
  doi={https://doi.org/10.1145/3544548.3581318}
}

@article{liang2025widespread,
  title={The widespread adoption of large language model-assisted writing across society},
  author={Liang, Weixin and Zhang, Yaohui and Codreanu, Mihai and Wang, Jiayu and Cao, Hancheng and Zou, James},
  journal={arXiv preprint arXiv:2502.09747},
  year={2025},
  doi={https://doi.org/10.48550/arXiv.2502.09747}
}

@article{westerlund2019emergence,
  title={The emergence of deepfake technology: A review},
  author={Westerlund, Mika},
  journal={Technology innovation management review},
  volume={9},
  number={11},
  year={2019},
  doi={http://doi.org/10.22215/timreview/1282}
}

@article{al2023impact,
  title={Impact of deepfake technology on social media: Detection, misinformation and societal implications},
  author={Al-Khazraji, Samer Hussain and Saleh, Hassan Hadi and Khalid, Adil Ibrahim and Mishkhal, Israa Adnan},
  journal={The Eurasia Proceedings of Science Technology Engineering and Mathematics},
  volume={23},
  number={429-441},
  pages={2},
  year={2023},
  publisher={ISRES Publishing},
  doi={https://doi.org/10.55549/epstem.1371792}
}

@inproceedings{lima2025public,
  title={Public Opinions About Copyright for AI-Generated Art: The Role of Egocentricity, Competition, and Experience},
  author={Lima, Gabriel and Grgi{\'c}-Hla{\v{c}}a, Nina and Redmiles, Elissa M},
  booktitle={Proceedings of the 2025 CHI Conference on Human Factors in Computing Systems},
  pages={1--32},
  year={2025},
  doi={https://doi.org/10.1145/3706598.3713338}
}

@inproceedings{kyi2025governance,
  title={Governance of Generative AI in Creative Work: Consent, Credit, Compensation, and Beyond},
  author={Kyi, Lin and Mahuli, Amruta and Silberman, M Six and Binns, Reuben and Zhao, Jun and Biega, Asia J},
  booktitle={Proceedings of the 2025 CHI Conference on Human Factors in Computing Systems},
  pages={1--16},
  year={2025},
  doi={https://doi.org/10.1145/3706598.3713799}
}

@article{samuelson2023generative,
  title={Generative AI meets copyright},
  author={Samuelson, Pamela},
  journal={Science},
  volume={381},
  number={6654},
  pages={158--161},
  year={2023},
  publisher={American Association for the Advancement of Science},
  doi={10.1126/science.adi0656}
}

@misc{pew2024socialmedia,
  title        = {Social Media Fact Sheet},
  author       = {{Pew Research Center}},
  year         = {2024},
  url          = {https://www.pewresearch.org/internet/fact-sheet/social-media/},
  note      = {Accessed: 2025-09-06},
}

@misc{mahdawi2025slop,
  title        = {AI‐generated ‘slop’ is slowly killing the internet, so why is nobody trying to stop it?},
  author       = {The Guardian},
  year         = {2025},
  url          = {https://www.theguardian.com/global/commentisfree/2025/jan/08/ai-generated-slop-slowly-killing-internet-nobody-trying-to-stop-it},
  note         = {Accessed: 2025-09-10},
}

@inproceedings{10.1145/3706598.3713522,
author = {He, Jessica and Houde, Stephanie and Weisz, Justin D.},
title = {Which Contributions Deserve Credit? Perceptions of Attribution in Human-AI Co-Creation},
year = {2025},
isbn = {9798400713941},
publisher = {Association for Computing Machinery},
address = {New York, NY, USA},
url = {https://doi.org/10.1145/3706598.3713522},
doi = {10.1145/3706598.3713522},
booktitle = {Proceedings of the 2025 CHI Conference on Human Factors in Computing Systems},
articleno = {540},
numpages = {18},
location = {
},
series = {CHI '25}
}

@inproceedings{gero2025creative,
  title={Creative Writers' Attitudes on Writing as Training Data for Large Language Models},
  author={Gero, Katy Ilonka and Desai, Meera and Schnitzler, Carly and Eom, Nayun and Cushman, Jack and Glassman, Elena L},
  booktitle={Proceedings of the 2025 CHI Conference on Human Factors in Computing Systems},
  pages={1--16},
  year={2025},
  doi={https://doi.org/10.1145/3706598.3713287}
}

@inproceedings{shan2023glaze,
  title={Glaze: Protecting artists from style mimicry by Text-to-Image models},
  author={Shan, Shawn and Cryan, Jenna and Wenger, Emily and Zheng, Haitao and Hanocka, Rana and Zhao, Ben Y},
  booktitle={32nd USENIX Security Symposium (USENIX Security 23)},
  pages={2187--2204},
  year={2023},
  url = {https://www.usenix.org/conference/usenixsecurity23/presentation/shan}
}

@inproceedings{shan2024nightshade,
  title={Nightshade: Prompt-specific poisoning attacks on text-to-image generative models},
  author={Shan, Shawn and Ding, Wenxin and Passananti, Josephine and Wu, Stanley and Zheng, Haitao and Zhao, Ben Y},
  booktitle={2024 IEEE Symposium on Security and Privacy (SP)},
  pages={807--825},
  year={2024},
  organization={IEEE},
  doi={10.1109/SP54263.2024.00207}
}

@article{jhaver2023personalizing,
  title={Personalizing content moderation on social media: User perspectives on moderation choices, interface design, and labor},
  author={Jhaver, Shagun and Zhang, Alice Qian and Chen, Quan Ze and Natarajan, Nikhila and Wang, Ruotong and Zhang, Amy X},
  journal={Proceedings of the ACM on Human-Computer Interaction},
  volume={7},
  number={CSCW2},
  pages={1--33},
  year={2023},
  publisher={ACM New York, NY, USA},
  doi={https://doi.org/10.1145/3610080}
}

@inproceedings{li2025beyond,
  title={Beyond Explicit and Implicit: How Users Provide Feedback to Shape Personalized Recommendation Content},
  author={Li, Wenqi and Kuo, Jui-Ching and Sheng, Manyu and Zhang, Pengyi and Wu, Qunfang},
  booktitle={Proceedings of the 2025 CHI Conference on Human Factors in Computing Systems},
  pages={1--17},
  year={2025},
  doi={https://doi.org/10.1145/3706598.3713241}
}

@inproceedings{ha2024organic,
  title={Organic or diffused: Can we distinguish human art from ai-generated images?},
  author={Ha, Anna Yoo Jeong and Passananti, Josephine and Bhaskar, Ronik and Shan, Shawn and Southen, Reid and Zheng, Haitao and Zhao, Ben Y},
  booktitle={Proceedings of the 2024 on ACM SIGSAC Conference on Computer and Communications Security},
  pages={4822--4836},
  year={2024},
  doi={https://doi.org/10.1145/3658644.3670306}
}

@article{boutadjine2025human,
  title={Human vs. machine: A comparative study on the detection of AI-generated content},
  author={Boutadjine, Amal and Harrag, Fouzi and Shaalan, Khaled},
  journal={ACM Transactions on Asian and Low-Resource Language Information Processing},
  volume={24},
  number={2},
  pages={1--26},
  year={2025},
  publisher={ACM New York, NY},
  doi={https://doi.org/10.1145/3708889}
}

@article{lundberg2025potential,
  title={The potential effects of deepfakes on news media and entertainment},
  author={Lundberg, Ebba and Mozelius, Peter},
  journal={AI \& SOCIETY},
  volume={40},
  number={4},
  pages={2159--2170},
  year={2025},
  publisher={Springer},
  doi={https://doi.org/10.1007/s00146-024-02072-1}
}

@inproceedings{gamage2022deepfakes,
  title={Are deepfakes concerning? analyzing conversations of deepfakes on reddit and exploring societal implications},
  author={Gamage, Dilrukshi and Ghasiya, Piyush and Bonagiri, Vamshi and Whiting, Mark E and Sasahara, Kazutoshi},
  booktitle={Proceedings of the 2022 CHI conference on human factors in computing systems},
  pages={1--19},
  year={2022},
  doi={https://doi.org/10.1145/3491102.3517446}
}

@inproceedings{habib2022okay,
  title={“Okay, whatever”: An evaluation of cookie consent interfaces},
  author={Habib, Hana and Li, Megan and Young, Ellie and Cranor, Lorrie},
  booktitle={Proceedings of the 2022 CHI conference on human factors in computing systems},
  pages={1--27},
  year={2022},
  doi={https://doi.org/10.1145/3491102.3501985}
}

@article{10.1145/3359174,
author = {McDonald, Nora and Schoenebeck, Sarita and Forte, Andrea},
title = {Reliability and Inter-rater Reliability in Qualitative Research: Norms and Guidelines for CSCW and HCI Practice},
year = {2019},
issue_date = {November 2019},
publisher = {Association for Computing Machinery},
address = {New York, NY, USA},
volume = {3},
number = {CSCW},
url = {https://doi.org/10.1145/3359174},
doi = {10.1145/3359174},
journal = {Proc. ACM Hum.-Comput. Interact.},
month = nov,
articleno = {72},
numpages = {23}
}

@article{armstrong1997place,
  title={The place of inter-rater reliability in qualitative research: An empirical study},
  author={Armstrong, David and Gosling, Ann and Weinman, John and Marteau, Theresa},
  journal={Sociology},
  volume={31},
  number={3},
  pages={597--606},
  year={1997},
  publisher={Cambridge University Press},
  doi={https://doi.org/10.1177/0038038597031003015}
}

@inproceedings{jiang2023evading,
  title={Evading watermark based detection of ai-generated content},
  author={Jiang, Zhengyuan and Zhang, Jinghuai and Gong, Neil Zhenqiang},
  booktitle={Proceedings of the 2023 ACM SIGSAC Conference on Computer and Communications Security},
  pages={1168--1181},
  year={2023},
  doi={https://doi.org/10.1145/3576915.3623189}
}

@misc{mantzarlis_dutta_2025,
  title        = {Tech platforms promised to label AI content. They're not delivering},
  author       = {Indicator},
  url = {https://indicator.media/p/tech-platforms-fail-to-label-ai-content-c2pa-metadata},
  year         = {2025},
  note         = {Accessed: 2025-11-17}
}

@misc{spotify_2025,
  title        = {Spotify to label AI music, filter spam and more in AI policy change},
  author       = {TechCrunch},
  url = {https://techcrunch.com/2025/09/25/spotify-updates-ai-policy-to-label-tracks-cut-down-on-spam/},
  year         = {2025},
  note         = {Accessed: 2025-11-17}
}

@article{are2025content,
  title={Content creators’ hopes and fears about artificial intelligence},
  author={Are, Carolina and Briggs, Pam and Brown, Richard},
  journal={Convergence},
  pages={13548565251372830},
  year={2025},
  publisher={SAGE Publications Sage UK: London, England},
  doi={https://doi.org/10.1177/13548565251372830}
}

@article{lao2025everyday,
  title={Everyday encounters with deepfakes: young people’s media and information literacy practices with AI-generated media},
  author={Lao, Yucong and Hirvonen, Noora and Larsson, Stefan},
  journal={Journal of Documentation},
  volume={81},
  number={7},
  pages={216--235},
  year={2025},
  publisher={Emerald Publishing Limited},
  doi={https://doi.org/10.1108/JD-01-2025-0007}
}

@article{pawelec2025decent,
  title={Decent deepfakes? Professional deepfake developers’ ethical considerations and their governance potential},
  author={Pawelec, Maria},
  journal={AI and Ethics},
  volume={5},
  number={3},
  pages={2641--2666},
  year={2025},
  publisher={Springer},
  doi={https://doi.org/10.1007/s43681-024-00542-2}
}

@inproceedings{ruth2025first,
  title={A first look at governments' enterprise security guidance},
  author={Ruth, Kimberly and Obu, Raymond Buernor and Shode, Ifeoluwa and Li, Gavin and Gates, Carrie and Ho, Grant and Durumeric, Zakir},
  booktitle={34th USENIX Security Symposium (USENIX Security 25)},
  pages={119--138},
  year={2025},
  url={https://www.usenix.org/conference/usenixsecurity25/presentation/ruth}
}

@misc{eu_dsa,
  author={EU Digital Strategy},
  title        = {The Digital Services Act},
  url = {https://digital-strategy.ec.europa.eu/en/policies/digital-services-act},
  note         = {Accessed: 2026-01-24},
  year         = {2022}
}

@inproceedings{10.1145/3715336.3735683,
author = {Huang, Jeff and Yew, Rui-Jie and Venkatasubramanian, Suresh},
title = {Copyrighting Generative AI Co-Creations},
year = {2025},
isbn = {9798400714856},
publisher = {Association for Computing Machinery},
address = {New York, NY, USA},
url = {https://doi.org/10.1145/3715336.3735683},
doi = {10.1145/3715336.3735683},
booktitle = {Proceedings of the 2025 ACM Designing Interactive Systems Conference},
pages = {1156–1164},
numpages = {9},
location = {
},
series = {DIS '25}
}

@inbook{thematicanalysis,
author = {Braun, Virginia and Clarke, Victoria},
year = {2012},
month = {01},
pages = {57-71},
title = {Thematic analysis.},
booktitle = {APA handbook of research methods in psychology 2},
editor = {Rindskopf D. and Sher K.J.},
isbn = {978-1-4338-1003-9},
publisher = {American Psychological Association},
}

@book{braun2021thematic,
  title={Thematic analysis: A practical guide},
  author={Braun, Virginia and Clarke, Victoria},
  year={2021},
  publisher={SAGE publications Ltd}
}

@inproceedings{prollochs2022community,
  title={Community-based fact-checking on Twitter’s Birdwatch platform},
  author={Pr{\"o}llochs, Nicolas},
  booktitle={Proceedings of the International AAAI Conference on Web and Social Media},
  volume={16},
  pages={794--805},
  year={2022},
  doi={https://doi.org/10.1609/icwsm.v16i1.19335}
}

@inproceedings{luna2024navigating,
  title={Navigating governance paradigms: A cross-regional comparative study of generative ai governance processes \& principles},
  author={Luna, Jose and Tan, Ivan and Xie, Xiaofei and Jiang, Lingxiao},
  booktitle={Proceedings of the AAAI/ACM Conference on AI, Ethics, and Society},
  volume={7},
  pages={917--931},
  year={2024},
  doi={https://doi.org/10.1609/aies.v7i1.31692}
}

@inproceedings{kolla2024llm,
  title={Llm-mod: Can large language models assist content moderation?},
  author={Kolla, Mahi and Salunkhe, Siddharth and Chandrasekharan, Eshwar and Saha, Koustuv},
  booktitle={Extended Abstracts of the CHI Conference on Human Factors in Computing Systems},
  pages={1--8},
  year={2024},
  doi={https://doi.org/10.1145/3613905.3650828}
}

@inproceedings{thomas2025supporting,
  title={Supporting Human Raters with the Detection of Harmful Content using Large Language Models},
  author={Thomas, Kurt and Kelley, Patrick Gage and Tao, David and Meiklejohn, Sarah and Vallis, Owen and Tan, Shunwen and Bratani{\v{c}}, Bla{\v{z}} and Ferreira, Felipe Tiengo and Eranti, Vijay Kumar and Bursztein, Elie},
  booktitle={2025 IEEE Symposium on Security and Privacy (SP)},
  pages={2772--2789},
  year={2025},
  organization={IEEE},
  doi={10.1109/SP61157.2025.00082}
}

@misc{communitynotes_x_api_overview,
  author       = {X Developer Documentation},
  title        = {AI Note Writers API Overview},
  url = {https://communitynotes.x.com/guide/en/api/overview},
  note         = {Accessed: 2026-01-24},
  year         = {2026}
}

\appendix
\onecolumn
\section{Appendix}
\subsection{Final Codebook}
\begin{table}[H]
\caption{Final codebook (Part 1), showing sub-codes and their sub-codes under Parent Code \textsc{AI-Generated Content Governance}}
\label{tab:finalcode1}
\renewcommand{\arraystretch}{1.2}
\begin{tabular}{p{0.23\textwidth}|p{0.67\textwidth}}
\toprule
\textbf{Sub-Code}          & \textbf{Description}                                                                                   \\
\midrule
\textsc{Governance Rationale}           & \textit{Why platforms govern AI-generated content posted and shared by users}                                   \\
\midrule
\textsc{User Rules}                      &               \textit{Rules on creating, posting, and sharing AI-generated content}\\                                                                                         \\
Rule                           & Rules on posting and sharing AI-generated content                                                      \\
Definition                     & Definitions on different kinds of AI-generated content                                                 \\
Example                        & Examples on the allowed or restricted AI-generated content                                             \\
\midrule
\textsc{Detection Methods}              &               \textit{How platforms identify AI-generated content and rule violations} \\                                                                                         \\
Violation detection            & Platforms identify AI-generated content that violates rules                                            \\
AI-generated content detection & Platforms detect if content is AI-generated                                                            \\
User report                    & Platforms enable users to report on AI-generated content, including content that violates rules                   \\
\midrule
\textsc{Governance Consequences}        &                             \textit{Enforcement on AI-generated content rules and user appeals} \\                                                                           \\
Enforcement on user            & Platforms punish users who post and share AI-generated content that violates rules                     \\
Enforcement on content         & Platforms moderate AI-generated content, including that violates rules                                   \\
Appeal/redress                 & Platforms enable (or do not enable) users to appeal the governance consequences applied to their content or themselves \\
\bottomrule
\end{tabular}
\end{table}

\begin{table}[H]
\caption{Final codebook (Part 2), showing sub-codes and their sub-codes under Parent Code \textsc{Integrated AI Tool Governance}}
\label{tab:finalcode2}
\renewcommand{\arraystretch}{1.2}
\begin{tabular}{p{0.23\textwidth}|p{0.67\textwidth}}
\toprule
\textbf{Sub-Code}          & \textbf{Description}                                                                                   \\
\midrule
\textsc{Governance Rationale}           & \textit{Why platforms govern integrated AI tools}                                   \\
\midrule
\textsc{User Rules}                      &               \textit{Rules on creating, posting, and sharing AI-generated content} \\                                                                                         \\
AI terms                           & Separated terms of use for integrated AI tools                                                      \\
Tool use                     & Rules on creating AI-generated content using integrated AI tools                                                 \\
Output distribution                        & Rules on sharing output from platform's integrated AI tools                            \\
\midrule
\textsc{Detection Methods}              &                          \textit{How platforms identify AI-generated content and rule violations} \\                                                                              \\
Tool use            & Platforms detect or prevent output generation that violate rules                                            \\
Output distribution & Platforms detect or prevent output distributions that violate rules                                                            \\
\midrule
\textsc{Governance Consequences}        &                    \textit{Enforcement on AI-generated content rules and user appeals} \\                                                                                   \\
Tool use           & Platforms deal with input, output, and users that violate rules when using integrated AI tools                     \\
Output distribution         & Platforms manage distributed output to ensure  trust, safety, and rule compliance                                   \\
\bottomrule
\end{tabular}
\end{table}

\begin{table}[H]
\caption{Final codebook (Part 3), showing sub-codes under Parent Code \textsc{User Empowerment}}
\label{tab:finalcode3}
\renewcommand{\arraystretch}{1.2}
\begin{tabular}{p{0.23\textwidth}|p{0.67\textwidth}}
\toprule
\textbf{Sub-Code}     & \textbf{Description}                                                                   \\
\midrule
AI labels             & Labeling system and disclosure labels for AI-generated content                          \\
User controls         & Recommendation controls and personalized filters on AI-generated content               \\
Educational materials & Materials that educate users on how to critically post, share, and identify AI-generated content\\
\bottomrule
\end{tabular}
\end{table}
\subsection{Dataset Overview}
\begin{table}[H]
\centering
\caption{Presence of governance and features of AI-generated content, across the 40 social media platforms we studied. ``Integrated AI Tool'' indicates a platform offers an integrated AI tool, ``Platform AI-Generated Content'' indicates that the platform uses AI to generate content for the community, and ``Governance'' indicates that the platform has enacted governance actions over AI-generated content.}
\label{tab:metastats1}
\renewcommand{\arraystretch}{0.95}
\setlength{\tabcolsep}{6pt}
\resizebox{0.65\textwidth}{!}{
\begin{tabular}{lccc}
\toprule
\multirow{2}{*}{\textbf{Platform}} & \multicolumn{2}{c}{\textbf{Features}}                       & \multirow{2}{*}{\textbf{Governance}} \\ \cline{2-3}
                          & \textit{Integrated AI Tool} & \textit{Platform AI-Generated Content} &                             \\
\midrule

Behance      & \textcolor{ForestGreen}{\ding{51}} & \textcolor{gray!70}{\ding{55}} & \textcolor{ForestGreen}{\ding{51}} \\
Bluesky      & \textcolor{gray!70}{\ding{55}}      & \textcolor{gray!70}{\ding{55}} & \textcolor{gray!70}{\ding{55}} \\
Dailymotion  & \textcolor{gray!70}{\ding{55}}      & \textcolor{gray!70}{\ding{55}} & \textcolor{ForestGreen}{\ding{51}} \\
DeviantArt   & \textcolor{ForestGreen}{\ding{51}} & \textcolor{gray!70}{\ding{55}} & \textcolor{ForestGreen}{\ding{51}} \\
Eporner      & \textcolor{gray!70}{\ding{55}}      & \textcolor{gray!70}{\ding{55}} & \textcolor{gray!70}{\ding{55}} \\
Erome        & \textcolor{gray!70}{\ding{55}}      & \textcolor{gray!70}{\ding{55}} & \textcolor{gray!70}{\ding{55}} \\

Facebook     & \textcolor{ForestGreen}{\ding{51}} & \textcolor{gray!70}{\ding{55}} & \textcolor{ForestGreen}{\ding{51}} \\
Fandom       & \textcolor{gray!70}{\ding{55}}      & \textcolor{ForestGreen}{\ding{51}} & \textcolor{gray!70}{\ding{55}} \\
Flickr       & \textcolor{gray!70}{\ding{55}}      & \textcolor{gray!70}{\ding{55}} & \textcolor{ForestGreen}{\ding{51}} \\
Goodreads    & \textcolor{gray!70}{\ding{55}}      & \textcolor{gray!70}{\ding{55}} & \textcolor{gray!70}{\ding{55}} \\

Imgur        & \textcolor{gray!70}{\ding{55}}      & \textcolor{gray!70}{\ding{55}} & \textcolor{gray!70}{\ding{55}} \\
Instagram    & \textcolor{ForestGreen}{\ding{51}} & \textcolor{gray!70}{\ding{55}} & \textcolor{ForestGreen}{\ding{51}} \\
LinkedIn     & \textcolor{ForestGreen}{\ding{51}} & \textcolor{ForestGreen}{\ding{51}} & \textcolor{ForestGreen}{\ding{51}} \\
LiveJournal  & \textcolor{gray!70}{\ding{55}}      & \textcolor{gray!70}{\ding{55}} & \textcolor{gray!70}{\ding{55}} \\
Medium       & \textcolor{gray!70}{\ding{55}}      & \textcolor{ForestGreen}{\ding{51}} & \textcolor{ForestGreen}{\ding{51}} \\

OnlyFans     & \textcolor{gray!70}{\ding{55}}      & \textcolor{gray!70}{\ding{55}} & \textcolor{ForestGreen}{\ding{51}} \\
Patreon      & \textcolor{gray!70}{\ding{55}}      & \textcolor{gray!70}{\ding{55}} & \textcolor{ForestGreen}{\ding{51}} \\
Pinterest    & \textcolor{ForestGreen}{\ding{51}} & \textcolor{gray!70}{\ding{55}} & \textcolor{ForestGreen}{\ding{51}} \\
Pixabay      & \textcolor{gray!70}{\ding{55}}      & \textcolor{gray!70}{\ding{55}} & \textcolor{ForestGreen}{\ding{51}} \\
Pixiv        & \textcolor{gray!70}{\ding{55}}      & \textcolor{gray!70}{\ding{55}} & \textcolor{ForestGreen}{\ding{51}} \\
Pornhub        & \textcolor{gray!70}{\ding{55}} & \textcolor{gray!70}{\ding{55}} & \textcolor{ForestGreen}{\ding{51}} \\
Quora          & \textcolor{ForestGreen}{\ding{51}} & \textcolor{ForestGreen}{\ding{51}} & \textcolor{ForestGreen}{\ding{51}} \\
Reddit         & \textcolor{ForestGreen}{\ding{51}} & \textcolor{ForestGreen}{\ding{51}} & \textcolor{ForestGreen}{\ding{51}} \\
ResearchGate   & \textcolor{gray!70}{\ding{55}} & \textcolor{gray!70}{\ding{55}} & \textcolor{gray!70}{\ding{55}} \\
Roblox         & \textcolor{ForestGreen}{\ding{51}} & \textcolor{gray!70}{\ding{55}} & \textcolor{ForestGreen}{\ding{51}} \\
SoundCloud     & \textcolor{ForestGreen}{\ding{51}} & \textcolor{gray!70}{\ding{55}} & \textcolor{ForestGreen}{\ding{51}} \\
StackOverflow  & \textcolor{ForestGreen}{\ding{51}} & \textcolor{ForestGreen}{\ding{51}} & \textcolor{ForestGreen}{\ding{51}} \\
SteamCommunity & \textcolor{gray!70}{\ding{55}} & \textcolor{gray!70}{\ding{55}} & \textcolor{gray!70}{\ding{55}} \\
Substack       & \textcolor{ForestGreen}{\ding{51}} & \textcolor{gray!70}{\ding{55}} & \textcolor{gray!70}{\ding{55}} \\
Threads        & \textcolor{ForestGreen}{\ding{51}} & \textcolor{gray!70}{\ding{55}} & \textcolor{ForestGreen}{\ding{51}} \\
TikTok         & \textcolor{ForestGreen}{\ding{51}} & \textcolor{ForestGreen}{\ding{51}} & \textcolor{ForestGreen}{\ding{51}} \\
TradingView    & \textcolor{gray!70}{\ding{55}} & \textcolor{gray!70}{\ding{55}} & \textcolor{gray!70}{\ding{55}} \\
Tumblr         & \textcolor{gray!70}{\ding{55}} & \textcolor{gray!70}{\ding{55}} & \textcolor{gray!70}{\ding{55}} \\
Twitch         & \textcolor{gray!70}{\ding{55}} & \textcolor{gray!70}{\ding{55}} & \textcolor{ForestGreen}{\ding{51}} \\
Twitter/X      & \textcolor{ForestGreen}{\ding{51}} & \textcolor{gray!70}{\ding{55}} & \textcolor{ForestGreen}{\ding{51}} \\
Vimeo          & \textcolor{ForestGreen}{\ding{51}} & \textcolor{ForestGreen}{\ding{51}} & \textcolor{ForestGreen}{\ding{51}} \\
XHamster       & \textcolor{gray!70}{\ding{55}} & \textcolor{gray!70}{\ding{55}} & \textcolor{ForestGreen}{\ding{51}} \\
XVideos        & \textcolor{gray!70}{\ding{55}} & \textcolor{gray!70}{\ding{55}} & \textcolor{gray!70}{\ding{55}} \\
Yelp           & \textcolor{gray!70}{\ding{55}} & \textcolor{ForestGreen}{\ding{51}} & \textcolor{ForestGreen}{\ding{51}} \\
YouTube        & \textcolor{ForestGreen}{\ding{51}} & \textcolor{ForestGreen}{\ding{51}} & \textcolor{ForestGreen}{\ding{51}} \\
\bottomrule
\multicolumn{4}{l}{\footnotesize \textcolor{ForestGreen}{\ding{51}}: passage(s) about such topics exist in our dataset; \textcolor{gray!70}{\ding{55}}: passage(s) about such topics do not exist in our dataset. }
\end{tabular}}
\end{table}

\begin{table}[H]
\centering
\caption{Location of information around governance of AI-generated content, across 27 social media platforms stating such information.}
\label{tab:location}
\resizebox{\textwidth}{!}{
\renewcommand{\arraystretch}{1.15}
\setlength{\tabcolsep}{3pt}
\begin{tabular}{p{80pt}*{27}{p{13pt}}}
\toprule
\rule{0pt}{1.8cm}&
\rothead{Behance} & \rothead{Dailymotion} & \rothead{DeviantArt} & \rothead{Facebook} &
\rothead{Flickr} & \rothead{Instagram} & \rothead{LinkedIn} & \rothead{Medium} &
\rothead{OnlyFans} & \rothead{Patreon} & \rothead{Pinterest} & \rothead{Pixabay} &
\rothead{Pixiv} & \rothead{Pornhub} & \rothead{Quora} & \rothead{Reddit} & \rothead{Roblox}& \rothead{SoundCloud} &
\rothead{StackOverflow} & \rothead{Threads} & \rothead{TikTok} & \rothead{Twitch} &
\rothead{Twitter/X} & \rothead{Vimeo} & \rothead{XHamster} & \rothead{Yelp} & \rothead{YouTube} \\
\midrule
Legal/Policy
  & \textcolor{gray!70}{\ding{55}} 
  & \textcolor{ForestGreen}{\ding{51}} 
  & \textcolor{ForestGreen}{\ding{51}} 
  & \textcolor{ForestGreen}{\ding{51}} 
  & \textcolor{ForestGreen}{\ding{51}} 
  & \textcolor{ForestGreen}{\ding{51}} 
  & \textcolor{ForestGreen}{\ding{51}} 
  & \textcolor{ForestGreen}{\ding{51}} 
  & \textcolor{ForestGreen}{\ding{51}} 
  & \textcolor{ForestGreen}{\ding{51}} 
  & \textcolor{ForestGreen}{\ding{51}} 
  & \textcolor{ForestGreen}{\ding{51}} 
  & \textcolor{ForestGreen}{\ding{51}} 
  & \textcolor{ForestGreen}{\ding{51}} 
  & \textcolor{ForestGreen}{\ding{51}} 
  & \textcolor{ForestGreen}{\ding{51}} 
  & \textcolor{ForestGreen}{\ding{51}} 
  & \textcolor{ForestGreen}{\ding{51}} 
  & \textcolor{ForestGreen}{\ding{51}} 
  & \textcolor{ForestGreen}{\ding{51}} 
  & \textcolor{ForestGreen}{\ding{51}} 
  & \textcolor{ForestGreen}{\ding{51}} 
  & \textcolor{ForestGreen}{\ding{51}} 
  & \textcolor{ForestGreen}{\ding{51}} 
  & \textcolor{ForestGreen}{\ding{51}} 
  & \textcolor{ForestGreen}{\ding{51}}    
  & \textcolor{ForestGreen}{\ding{51}} 
\\
Support/Help Center
  & \textcolor{ForestGreen}{\ding{51}} 
  & \textcolor{gray!70}{\ding{55}}    
  & \textcolor{ForestGreen}{\ding{51}} 
  & \textcolor{ForestGreen}{\ding{51}} 
  & \textcolor{ForestGreen}{\ding{51}} 
  & \textcolor{ForestGreen}{\ding{51}} 
  & \textcolor{ForestGreen}{\ding{51}} 
  & \textcolor{ForestGreen}{\ding{51}} 
  & \textcolor{ForestGreen}{\ding{51}} 
  & \textcolor{ForestGreen}{\ding{51}}    
  & \textcolor{ForestGreen}{\ding{51}} 
  & \textcolor{ForestGreen}{\ding{51}} 
  & \textcolor{ForestGreen}{\ding{51}} 
  & \textcolor{ForestGreen}{\ding{51}}    
  & \textcolor{gray!70}{\ding{55}}
  & \textcolor{ForestGreen}{\ding{51}} 
  & \textcolor{ForestGreen}{\ding{51}}
  & \textcolor{ForestGreen}{\ding{51}} 
  & \textcolor{ForestGreen}{\ding{51}} 
  & \textcolor{ForestGreen}{\ding{51}} 
  & \textcolor{ForestGreen}{\ding{51}} 
  & \textcolor{ForestGreen}{\ding{51}}    
  & \textcolor{ForestGreen}{\ding{51}} 
  & \textcolor{ForestGreen}{\ding{51}} 
  & \textcolor{gray!70}{\ding{55}} 
  & \textcolor{ForestGreen}{\ding{51}} 
  & \textcolor{ForestGreen}{\ding{51}} 
\\
News/Announcement
  & \textcolor{gray!70}{\ding{55}} 
  & \textcolor{gray!70}{\ding{55}}    
  & \textcolor{ForestGreen}{\ding{51}} 
  & \textcolor{ForestGreen}{\ding{51}} 
  & \textcolor{gray!70}{\ding{55}} 
  & \textcolor{ForestGreen}{\ding{51}} 
  & \textcolor{ForestGreen}{\ding{51}} 
  & \textcolor{ForestGreen}{\ding{51}} 
  & \textcolor{gray!70}{\ding{55}} 
  & \textcolor{ForestGreen}{\ding{51}}    
  & \textcolor{ForestGreen}{\ding{51}} 
  & \textcolor{ForestGreen}{\ding{51}} 
  & \textcolor{gray!70}{\ding{55}} 
  & \textcolor{gray!70}{\ding{55}}    
  & \textcolor{ForestGreen}{\ding{51}}
  & \textcolor{ForestGreen}{\ding{51}} 
  & \textcolor{ForestGreen}{\ding{51}}
  & \textcolor{ForestGreen}{\ding{51}} 
  & \textcolor{ForestGreen}{\ding{51}} 
  & \textcolor{ForestGreen}{\ding{51}} 
  & \textcolor{ForestGreen}{\ding{51}} 
  & \textcolor{ForestGreen}{\ding{51}}    
  & \textcolor{ForestGreen}{\ding{51}} 
  & \textcolor{ForestGreen}{\ding{51}} 
  & \textcolor{gray!70}{\ding{55}} 
  & \textcolor{ForestGreen}{\ding{51}} 
  & \textcolor{ForestGreen}{\ding{51}} 
\\
\bottomrule
\multicolumn{28}{p{1.15\textwidth}}{\textcolor{ForestGreen}{\ding{51}}: this type of page exists in our dataset, \textcolor{gray!70}{\ding{55}}: this type of page does not exist in our dataset. }
\end{tabular}
}
\end{table}









\end{document}